\documentclass[11pt]{article}
\usepackage{journals}      % AAS journal definitions
\usepackage{emulateapj}
\usepackage{flushrt}
\usepackage{natbib}
\usepackage{epsfig}

\newcommand{\HII}      {H{\footnotesize \,II} }
\newcommand{\Ncl}      {{\cal N}_{\rm cl}}
\newcommand{\Nstar}    {{\cal N}_{\star} }

\newcommand{\tmsbar}   {\left<t_{\rm ms}\right>}
\newcommand{\mbarstar} {\left<m_{\star}\right>}
\newcommand{\sbar}     {\left<s_{49}\right>}
\newcommand{\SFEtot}   {{\rm SFE_{\rm tot}}}

\newcommand{\rmerge}   {r_{m}}
\newcommand{\cII}      {c_{\rm II} }
\newcommand{\rII}      {r_{\rm II} }
\newcommand{\rhoII}      {\rho_{\rm II} }
\newcommand{\phiII}      {\phi_{\rm II} }
\newcommand{\Mclsix}	{M_{\rm cl,6}}

\newcommand{\pc}	{{\rm pc}}
\newcommand{\yr}	{{\rm yr}}
\newcommand{\Mcl}	{M_{\rm cl}}
\newcommand{\Rcl}	{R_{\rm cl}}
\newcommand{\Myr}	{{\rm Myr}}

\newcommand{\Msun}	{M_\odot}
\newcommand{\vrms}	{v_{\rm rms}}
\newcommand{\ceff}	{c_{\rm eff}}
\newcommand{\vesc}	{v_{\rm esc}}
\newcommand{\cm}	{{\rm cm}}
\newcommand{\tff}	{t_{\rm ff}}
\newcommand{\kps}	{{\rm km~s^{-1}}}
\begin{document}

% %------------------------------------------------------------------------
\title{On the Role of Massive Stars in the Support and Destruction of Giant
Molecular Clouds}
\author{Christopher D. Matzner} 
\affil{Canadian Institute for Theoretical Astrophysics, University of Toronto}
\begin{abstract} 
We argue that massive stars are the dominant sources of energy for the
turbulent motions within giant molecular clouds, and that the primary
agent of feedback is the expansion of \HII regions within the cloud
volume.  This conclusion is suggested by the low efficiency of star
formation and corroborated by dynamical models of \HII regions.  We
evaluate the turbulent energy input rate in clouds more massive than
$3.7\times 10^5$ solar masses, for which gravity does not
significantly affect the expansion of \HII regions.  Such clouds
achieve a balance between the decay of turbulent energy and its
regeneration in \HII regions; summed over clouds, the implied ionizing
luminosity and star formation rate are roughly consistent with the
Galactic total.  \HII regions also photoevaporate their clouds: we
derive cloud destruction times somewhat shorter than those estimated
by Williams and McKee.  The upper mass limit for molecular clouds in
the Milky Way may derive from the fact that larger clouds would
destroy themselves in less than one crossing time.  The conditions
within starburst galaxies do not permit giant molecular clouds to be
supported or destroyed by \HII regions.  This should lead to rapid
cloud collapse and the efficient formation of massive star clusters,
which may explain some aspects of the starburst phenomenon.
\end{abstract}

\section{Introduction}\label{Intro}
Giant molecular clouds (GMCs) are the sites of most star formation in
the Milky Way, and the evolution of the Galaxy and its stellar
population are controlled in large part by the physics of GMCs.
Despite several decades of observations, the clouds' physical nature,
longevity, and modes of formation and destruction are still matters of
debate.  If clouds survive for more than a single dynamical time, it
is plausible that their observed properties are (like stars') the
product of internal sources and sinks of energy \citep[for a recent
review, see][]{Crete99}. If instead they disappear after one crossing
time \citep{2000ApJ...530..277E} then their properties will be more
affected by the mechanism of their formation
\citep{1997ApJ...474..292V,1999ApJ...527..285B, 1999ApJ...515..286B}.
However, rapid destruction also implies that the means of cloud
destruction may play an equally important role. This paper considers
what consequences the mechanisms of cloud destruction -- primarily
\HII regions -- imply for the internal motions, energy budgets, star
formation rates, and lifetimes of GMCs.

\subsection{Cloud Scaling Laws and Their Origins}\label{SS:scaling}
Molecular clouds are observed to obey a set of scaling relations known
collectively as Larson's (\citeyear{1981MNRAS.194..809L}) laws.  As
updated by \cite{1987ApJ...319..730S}, these can be summarized: a
constant mean column density $\Sigma\simeq 170 \Msun/\pc^2$,
corresponding to a visual extinction $A_V\simeq 7.5$ mag; and virial
balance, with a virial parameter $\alpha \equiv 10\vrms^2/(3\vesc^2)$
\citep{BM92} of order unity \citep{1988ApJ...329..392M}.
\citeauthor{1987ApJ...319..730S} adopt the value $\alpha =1.11$,
yielding the relation $\vrms \simeq \vesc/\sqrt{3}$ between the
r.m.s. and escape velocities.\footnote{We only consider clouds like
those in the \citeauthor{1987ApJ...319..730S} survey, and assume this
survey correctly infers the clouds' properties.  As we shall concentrate on
the most massive clouds ($\Mcl > 3.7\times 10^5 \Msun$), whose
observations may be affected by beam smearing and velocity crowding
(J.~P. Williams 2001, private communication), this assumption should
be checked by future observations.}  A third but not independent
property is the scaling of line width ($\sigma$) with cloud radius
($\Rcl =R_{\rm pc}$ pc): $\sigma\equiv \vrms/\sqrt{3} = 0.72 R_{\rm
pc}^{1/2}$ km/s.  As they do not represent infall, these motions are
considered turbulence.  The molecular gas is cold ($\sim 10-30$ K) and
highly magnetized ($\sim 30 \mu{\rm G}$), so its motions are
supersonic but roughly Alfv\'enic.

\cite{1997ApJ...474..292V} advance a model in which these properties
derive from motions within the interstellar medium (ISM) from which
GMCs form.  In two-dimensional simulations of turbulence within the
Galactic disk, they identify a population of overdensities that could
be considered clouds.  Because of the turbulent spectrum, all of the
objects formed in their simulations obey the line width-size relation noted above.
However, all but the few most massive of these are transient
compressions rather than self-gravitating objects; hence their escape
velocities and surface densities do not follow the virial relation.
\citeauthor{1997ApJ...474..292V} suggest that selection effects
restrict observed clouds to a narrow range of inferred column
densities.  Specifically, they note that the IRAS survey of
\cite{1994ApJS...95..457W} may only be sensitive to an outer shell of
warm dust around clouds of various columns.  However, \cite{Crete99}
has countered that CO would form and be detectable at significantly
smaller columns \citep[$A_V=1.8$;][]{1988ApJ...334..771V} than are
typical of the observed GMCs. 

An alternative possibility is that the common GMC column density
arises from internal cloud processes: specifically energetic feedback
due to star formation \citep[as in the model of][hereafter M89]{M89}.
For this to be possible, two conditions must hold: 1. stars form
rapidly only in regions that exceed a critical column density; and
2. star formation is potentially so vigorous a source of turbulent
energy that it overwhelms the natural decay of turbulence if the
column is much above this critical value.  Under these conditions a
GMC will settle into a state of energetic equilibrium much like a
star's, with star formation occurring just fast enough to offset
turbulent decay. The necessary column density would be roughly the
critical value, hence the common value observed among GMCs.

For condition (1), \citeauthor{M89} proposed that star formation is
inhibited in regions that are not shielded by $\sim 4$ visual
magnitudes in each direction, because these layers are penetrated by
far-ultraviolet (FUV) photons that elevate the level of ionization.
FUV ionization thus slows ambipolar diffusion, which M89 argued to be
a rate-limiting step for star formation because stellar mass regions
are too highly magnetized to collapse directly.  This sets the
critical column density at roughly 8 visual magnitudes (4 on each
side), close to the value of 7.5 observed in GMCs.  Corroborating this
hypothesis, \cite{1998ApJ...502..296O} find a sharp distinction among
substructures in the Taurus clouds between those with $A_V\gtrsim 7.5$
mag that are actively forming stars, and those with lower extinctions
that are not.  In the low-metallicity environment of the SMC,
\cite{1998ApJ...498..735P} verified that the clouds maintain $A_V \sim
7.5$ mag, although this required a higher column density than for
Milky Way clouds.

For condition (2), \citeauthor{M89} specified protostellar winds as
the agents of star formation feedback, following \cite{NS80} and
\cite{1983ApJ...273..243F}, who implicated main-sequence winds, and
\cite{LG82}, who realized the potential of their protostellar
counterparts.  Subsequent numerical simulations
\citep[e.g.,][]{1998PhRvL..80.2754M, 1998ApJ...508L..99S,
1999ApJ...524..169M,2001ApJ...546..980O} have indicated a much faster
decay of turbulence than \citeauthor{M89} assumed, calling into
question \citep[e.g.,][]{2001ApJ...551..743B} the notion that energy
injection from protostellar winds is vigorous enough to offset
turbulent decay.  Below, we show that \HII regions \citep[originally
considered by][]{1975ApJ...196L..77A} represent an additional,
inevitable, and more important source of turbulent energy for
molecular clouds.

\subsection{Molecular Cloud Destruction and the Inefficiency of Star
Formation} \label{SS:destruction} 

The rapid decay of turbulent energy should logically cause an equally
rapid contraction of GMCs, the end product of which is star formation.
Nevertheless, clouds make stars at a tiny fraction of the rate allowed
by direct gravitational collapse \citep{1974ARA&A..12..279Z}.  Star
formation is not only slow, but inefficient: the protostellar sources
observed by \cite{1979ApJS...41..743C} compose $\sim 10\%$ of their
surrounding clouds; \cite{1986ApJ...301..398M} surveyed 54 clouds,
estimating $\sim 2\%$ of their mass to be in stars, on the basis of
their \HII regions; and \cite{1997ApJ...476..166W} (hereafter, WM97)
argued that only $10\%$ of the mass of a GMC would ever become
stellar.  More recently, \cite{2000AJ....120.3139C} has estimated that
$1\%-9\%$ of the mass of several nearby molecular clouds is in
embedded stars.  The sluggishness and inefficiency of star formation
can only be consistent with the rapid decay of turbulent energy if
GMCs are destroyed more rapidly than they can convert themselves into
stars.

What process destroys molecular clouds?  \cite{1999ApJ...515..286B}
note that gravitationally unbound clouds formed by turbulent
compressions \citep[most of the objects seen by][]
{1997ApJ...474..292V} are easily disrupted in a single crossing time
by the flows that created them.  However they also show that bound
objects, once formed, continue to collapse rather than
re-expanding.  
(In these authors' simulations, such clouds are stabilized or
disrupted by a local heating included to represent the action
of massive star formation, a topic to be addressed in detail below.)
The \citeauthor{1987ApJ...319..730S}\ survey indicates
that molecular clouds are far too tightly bound to be destroyed by
turbulence in the interstellar medium: their hydrostatic pressures are
in excess of $3\times 10^5 k_B~\cm^{-3}$, whereas the bounding gas
pressure is only $2\times 10^4 k_B~\cm^{-3}$ \citep{Crete99}.  The ram
pressure of motions in the diffuse ISM ($n_H\sim 1~\cm^{-3}$, $\sigma
\sim 10~\kps$) is also $\sim 2\times 10^4 k_B~\cm^{-3}$, and thus
insufficient to disrupt a cloud.
Clouds contain a binding energy per volume that is roughly three
halves their hydrostatic pressure \citep[somewhat less, due to
magnetization;][]{Crete99}; this was radiated in the process of
formation and must be resupplied to unbind the cloud.  The most
plausible source for this energy is young and massive stars formed
within the cloud itself.

Several mechanisms have been considered in this regard.
\cite{2000ApJ...545..364M} have shown that small clouds can be
disrupted by protostellar outflows as they form low-mass stars;
however, they predict such mass loss not to be important for giant
clouds.  \cite{1983MNRAS.203.1011E} suggested that radiation pressure
might unbind clouds if their stellar populations became too luminous,
but neglected the reprocessing of radiation into far-infrared
wavelengths to which clouds are transparent
\citep[see][]{1996ApJ...462..874J}.  Molecules are destroyed by FUV
photons in photodissociation regions; however, the thermal velocities
of these regions are far below the escape velocities of GMCs.
Therefore, photodissociation regions are thought to form an atomic
layer around GMCs \citep{1988ApJ...334..771V} and have only been
suggested as a means of destruction for clouds that are dynamically
disrupted by a different process \citep{1979MNRAS.186...59W,
1997ApJ...476..144M, 2001astro.ph..8023H}, as
\cite{1996ApJ...464..247W} may have observed around the cold cloud
G216-2.5. 

\cite{1979MNRAS.186...59W} calculated the destructive effects of
massive stars due to the ejection of photoionized gas, finding $\sim
10^4~\Msun$ to be ejected for each blister-type \HII region.
Supernova explosions typically add $\sim 10\%$ and at most $40\%$ to
this amount \citep{1989A&A...216..207Y}, for \HII regions created by
single stars.  A few supernovae arise from B stars in the mass range
$8-12~\Msun$, which do not have appreciable \HII regions
\citep{1999ApJ...511..798C}; however, these require at least
$1.3\times 10^7$ years to evolve, and eject only $\lesssim 500~\Msun$
each if they explode inside their cloud (and much less otherwise;
\citealt{1985A&A...145...70T}).  The main-sequence winds of massive
stars emit a total energy comparable to supernovae.
\cite{1987ApJ...317..190M} and \cite{1984ApJ...278L.115M} have argued
that stellar wind bubbles are confined within \HII regions in the
context of a cloudy medium.  For these reasons, we concentrate on
photoionization as the primary cause of destruction for GMCs (as have
\citealt{1979MNRAS.186...59W}, \citealt{1980ApJ...238..148B},
\citealt{1989A&A...216..207Y}, \citealt{1994ApJ...436..795F}, and
WM97).  The dynamical effects of stellar winds and supernovae are
considered in \S \ref{Winds&SNe}. 

The fact that massive stars are responsible for unbinding molecular
clouds more rapidly than they can form stars implies that they supply
more energy than is dissipated in turbulence.  If a fraction of this
energy is incorporated into the motions of the remaining molecular
gas, it will sustain turbulence and slow the cloud's contraction.
In the following sections we consider the loss and regeneration
of turbulent motions (\S \ref{SourcesAndSinks}), the relative
importance of \HII regions and protostellar winds (\S \ref{Dominance}),
the dynamics of an individual association (\S \ref{OBmomentum}) and of
a population of associations (\S \ref{PperStar}), and the 
implications of an equilibrium between the decay of turbulence and
its regeneration in \HII regions for the ionizing luminosities and star formation rates (\S
\ref{SFR}) and lifetimes (\S \ref{Lifetimes}) of massive clouds.
These estimates are corrected for the interaction between \HII regions
in \S \ref{Porosity}.  

Finally, in \S \ref{Conclusions} we point out that the high-pressure
environments of starburst galaxies prevent molecular clouds from being
destroyed or even supported by photoionization.  In the absence of
other sources of energy, such clouds must collapse and form stars at a
high efficiency; this is a recipe for the efficient formation of
massive star clusters often observed in starbursts. 

\section{Sources and Sinks of Turbulent Motion}\label{SourcesAndSinks}
To consider the fates of GMCs, one must account for the gains and
losses of turbulent energy \citep{NS80}.  {M89} assumed a range
of timescales for the dissipation of turbulent energy, 
\begin{equation}\label{diss}
t_{\rm diss} \equiv - \frac {E_{\rm turb}}{\dot{E}_{\rm turb, diss}}, 
\end{equation}
ranging from three to ten times the free-fall time of the cloud.
Here, $E_{\rm turb}$ represents both kinetic and magnetic energy
associated with turbulent motions \citep{1995ApJ...439..779Z}.
Simulations by \cite{1998ApJ...508L..99S} show incomplete
equipartition between these components: for the case that most
resembles molecular clouds (magnetic pressure 100 times greater than
gas pressure),
\begin{equation}\label{equipart}
E_{\rm turb}= 0.78 \Mcl\vrms^2. 
\end{equation}
% this is close to the value $[(1+2/3)\times 1/2 = 0.83]\Mcl \vrms^2$
% expected if the magnetic perturbations are in equipartition with
% cross-field motions. 

\citeauthor{M89}'s assumption of a relatively long decay time scale
was based on the notion that the magnetic field should cushion gas
motions.  Testing this notion numerically, several groups
\citep{1998PhRvL..80.2754M, 1998ApJ...508L..99S,
1999ApJ...524..169M,2001ApJ...546..980O} have found the dissipation to
be much more rapid, especially if turbulence is driven on scales
smaller than the entire cloud.  For the same physical situation that
gives equation (\ref{equipart}), \citeauthor{1998ApJ...508L..99S} find
\begin{equation}\label{tdissNum} 
t_{\rm diss} = 0.83 \frac{\lambda_{\rm in}}{\vrms},
\end{equation}
where $\lambda_{\rm in}$ is the wavelength on which the turbulence is
stirred.   Combining this with \cite{1987ApJ...319..730S}'s virial
relation $\vrms = \vesc/\sqrt{3}$ gives $t_{\rm diss}/\tff \simeq 0.93
{\lambda_{\rm in}}/{R_{\rm cl}}$.  This dissipation time scale is
shorter than the range assumed by {M89} if the forcing scale for
turbulence is smaller than the radius of the cloud; the importance of
the forcing scale has recently been highlighted by
\cite{2001ApJ...551..743B}.

To estimate how turbulence is driven one must allow for the radiative
nature of the gas, which causes compressions to be very dissipative.
Our argument follows \cite{NS80} and {M89}.  An impulse (such as a
protostellar wind) whose momentum is $\delta p$ will cause a
disturbance that decelerates as it sweeps into the cloud.  Since
energy is radiated, a thin shell forms and conserves linear momentum
in each direction \citep[e.g.,][]{1999ApJ...526L.109M}; the kinetic
energy associated with the motion is $v\delta p/2$ when the velocity
is $v$.  This continues until $v$ has decelerated to a terminal
velocity that WM97 estimate to be the effective sound speed,
$\ceff\equiv \sqrt{P/\rho}$.  (Note that $\ceff$ far exceeds the
thermal sound speed, as $P$ includes the total hydrostatic pressure.)
At this point the swept-up shell thickens, stalls, and loses
coherence, rendering its energy to the turbulence.  The increase of
turbulent energy is therefore
\begin{equation}\label{regen}
\delta E_{\rm turb} = \frac{\phi}{2} \ceff \delta p. 
\end{equation} 
The efficiency coefficient $\phi$ is uncertain, and must be determined
by simulation; \cite{Crete99} suggests $\phi\simeq 1.6$ to account for
the energy stored in magnetic perturbations at the end of
deceleration.  Equation (\ref{regen}) implies that protostellar winds
and \HII regions generate turbulent energy in proportion to the rate at
which they impart momentum to the cloud. 

Equation (\ref{regen}) is essentially the same formula employed
previously by \cite{NS80}, {M89}, \cite{BM96}, \cite{Crete99},
\cite{1999sf99.proc..353M} and \cite{myphd}.  It assumes that energy
is injected in an explosive manner, so that the early stage of
momentum-conserving thin shell expansion (at speeds above $\ceff$) can
be separated from the later stage of turbulent dissipation (at speeds
of about $\vrms$).  It is only valid in cases where the kinetic energy
of relative gas motions is present in the center-of-mass frame of the
cloud; therefore, it does not apply to large-scale gravitational
fields.  For the same reason its validity is suspect if the force that
gives rise to $\delta p$ is applied over a large length scale compared
to the cloud radius.  However, we shall see in the subsequent sections 
that \HII regions are explosive events that input momentum on scales
smaller than $\Rcl$, at least for clouds in our mass range of interest. 

What are appropriate values of $\delta p$?  In the formation of a
low-mass star, a fraction of the material that accretes onto the
star-disk system is redirected into winds with a characteristic
velocity $\sim 200$ km/s.  An observational analysis by
\cite{Richeretal} implies a characteristic wind momentum of roughly
$50 ~\kps$ times the mass of the star that forms; however, this
analysis is quite uncertain.  Models exist in which the wind removes
anywhere from a tenth to half the mass flowing through the disk
\citep{1992ApJ...394..117P,1994ApJ...429..808N}.  We adopt one-sixth
as an intermediate value \citep[following][]{2000ApJ...545..364M}, so
that the wind mass is five times smaller than the star's mass.  With
this choice, each star of mass $m_\star$ generates a wind impulse
\begin{equation}\label{pw}
\delta p_w = \phi_w m_\star\times 40~\kps
\end{equation}
where $\phi_w$ represents our uncertainty.  In \S \ref{Winds&SNe} we
find that main-sequence and evolved stars contribute a comparable
momentum in the six million years after their formation; here, we
restrict our attention to the more impulsive protostellar winds. 

If $\dot{M}_\star$ is the total rate at which mass is converted into
stars in a cloud, equation (\ref{regen}) gives
\begin{equation}\label{pdotstar}
\dot{E}_{{\rm turb},w} = \frac{\phi\phi_w}{2} \dot{M}_\star(40~\kps)
\ceff
\end{equation}
as the rate of turbulence regeneration by protostellar winds. 

Next, consider cloud destruction at a rate $\dot{M}_{\rm dest}$ in a
sequence of discrete events (ejecting $\delta M_{\rm dest}$ per
event), no one of which completely destroys the cloud.  Each mass
ejection delivers an equal and opposite impulse to the cloud material
left behind, and each impulse increases cloud turbulence according to
equation (\ref{regen}).  Since the ejection velocity must exceed
$\vesc$ near the surface of the cloud,
\begin{equation}\label{pdotDestBound}
\delta {p}_{\rm dest} \gtrsim \delta {M}_{\rm dest} \vesc. 
\end{equation}
If photoionization is the primary means of destruction, then material
is ejected at about the thermal velocity of ionized gas, $\cII\simeq 10
~\kps$, and
\begin{equation}\label{pdotIonization}
\delta {p}_{\rm dest} = \phiII \,\delta{M}_{\rm dest}\cII, 
\end{equation}
where $\phiII$ is defined in analogy to $\phi_w$; by equation (\ref{regen}), 
\begin{equation}\label{EdotIonization}
\dot{E}_{\rm turb, dest} = \frac{\phi\phiII}{2} \dot{M}_{\rm dest}
\cII \ceff.
\end{equation}
An analysis of the generation of momentum in ionization fronts,
presented in section \ref{IFmomentum}, shows that $\phiII \geq 2$.

\subsection{Massive Stars Dominate the Energy Budgets of
GMCs}\label{Dominance} 

A comparison between equations (\ref{pdotstar}) and
(\ref{EdotIonization}) reveals the importance of massive star
formation in the support of GMCs.  Relevant to this comparison is the
net efficiency of star formation, $\SFEtot$, defined as the fraction
of cloud mass that will ever become stars:
\begin{equation}\label{SFE}
\SFEtot \equiv \frac{\int\dot{M}_\star
dt}{\int(\dot{M}_\star + \dot{M}_{\rm dest}) dt};
\end{equation}
the denominator equals the initial mass of the cloud (in the absence
of ongoing accretion from the ISM).   Thus
\begin{equation}\label{SFE2}
\left<\dot{M}_{\rm dest}\right> = (\SFEtot^{-1}-1) 
\left<\dot{M}_\star\right> 
\end{equation}
where the brackets are time averages.  $\SFEtot$ is estimated
observationally by the {\em current} fraction of mass in stars; the
two are comparable if the efficiency is low, if stars remain within
their clouds \citep{2000ApJ...545..364M}.  
% \cite{2000AJ....120.3139C}
% gives a range of stellar mass fractions between $1\%$ and $9\%$ for
% nearby clouds, where the dispersion arises partly from the difference
% between clouds and partly from assumptions about the ages of embedded
% stars.  
If we take $\SFEtot\sim 5\%$ as typical, then $95\%$ of the
cloud's mass will be disrupted by blister \HII regions rather than
achieving stardom.  Equations (\ref{pw}), (\ref{pdotIonization}), and
(\ref{SFE2}) imply that the momentum contributed by protostellar winds
is smaller than the contribution by \HII regions so long as
\begin{equation}\label{dominance}
\SFEtot < \frac{1}{4{\phi_w}/{\phiII}+1}\simeq 33\%.
\end{equation}
Clouds that make stars as inefficiently as Galactic GMCs derive more
turbulent energy from blister \HII regions {\em alone} than from
protostellar winds.  In \S \ref{PperStar} we shall show that \HII
regions would be more important than protostellar winds even if star
formation were efficient. 

\section{Effect of a Single Association}\label{OBmomentum}
\subsection{Momentum Generation by an \HII Region} \label{IFmomentum}
The dynamical phases of expansion of an \HII region have been
presented in numerous prior works
\citep[e.g.,][]{1979MNRAS.186...59W,1978ppim.book.....S}.  These
authors neglected the inertia of the shell of shocked cloud gas; we
give only a cursory treatment including this inertia, for the purpose
of quantifying the momentum generated.  We follow
\cite{1997ApJ...476..144M} (hereafter, MW97) in adopting a common
temperature of 7000 K for ionized gas (and mean molecular weight 0.61,
so $\cII = 9.74$ km/s) and in approximating that $27\%$ of the
ionizing photons (emitted at a rate $10^{49}S_{49}$ per second) are
absorbed by dust rather than gas.  We will also take for the molecular
gas a uniform density $\rho_0$ and hydrogen number density $n_H$
consistent with a mean hydrogen column density $N_{H,22} \equiv
N_H/(10^{22}\cm^{-2}) = 1.5$ \citep{1987ApJ...319..730S}.

After a rapid expansion to the initial Str\"omgen radius \citep[using
the recombination coefficient of][]{1995MNRAS.272...41S}
\begin{equation}\label{Rst0}
R_{\rm St,0} = 2.9 \left(\frac{N_{H,22}}{1.5}\right)^{-1}
S_{49}^{1/3} \Mclsix^{1/3}~\pc, 
\end{equation}
the \HII region is governed by the requirement that a very small
fraction of the ionizing photons actually reach the ionization front.
This causes ionized gas density $\rho_{\rm II}$ to vary as
\begin{equation}\label{rhoII}
\frac{\rho_{\rm II}}{\rho_0} =\left( \frac{\rII}{R_{\rm
St,0}}\right)^{-3/2}, 
\end{equation}
where $r_{\rm II}$ is the current radius of the ionization front.  As
the ionization front expands subsonically the density $\rhoII$ and
pressure $\rhoII \cII^2$ are nearly uniform within the \HII region
(albeit more perfectly for embedded than for blister regions, in which
a pressure gradient develops as gas accelerates away). 

So long as the \HII region expands supersonically with respect to the
molecular gas, it is bounded by a thin shell of dense, shocked cloud
gas.  The radius of this shell is nearly identical to $\rII$, and we
shall restrict our attention to the period of expansion $\rII\gg
R_{St,0}$, when nearly all of the mass originally within $\rII$
remains within the shell \citep{1978ppim.book.....S}.  Let $A_{\rm
sh}$ denote the shell's area and $M_{\rm sh}$ its mass.  The shell's
momentum equation is
\begin{equation}\label{shellmom}
\frac{d}{dt} \left(M_{\rm sh} \dot{r}_{\rm II}\right) = A_{\rm sh}
\rhoII\left[ \cII^2 + u_{\rm II} (u_{\rm II} - \dot{r}_{\rm
II})\right].
\end{equation}
On the left of this equation is $d\delta{p}/dt$, the rate of increase of
the shell's momentum; on the right, the forces due to pressure (first
term in brackets) and due to thrust caused by the exhaust of ionized
gas at a velocity $u_{\rm II}$ relative to the cloud. 

Blister and embedded \HII regions differ in the coefficient relating
$A_{\rm sh}$ to $\rII$: $A_{\rm sh} = (1,2)\times 2 \pi \rII^2$ for
(blister, embedded) HII regions whose ionization fronts are idealized
as hemispheres and spheres, respectively.

Another difference is the relative importance of thrust and pressure
in generating momentum.  In an embedded region, the ionized gas is
trapped and only expands homologously; equation (\ref{rhoII}) then
implies $u_{\rm II} = \dot{r}_{\rm II}/2$, which is much less than
$\cII$ when $\rII\gg R_{\rm St, 0}$ \citep{1978ppim.book.....S} so
that only pressure need be considered.  In blister regions, on the
other hand, ionized gas flows away freely allowing the ionization
front to tend toward the D-critical state \citep{1954BAN....12..187K}
for which $u_{\rm II}-\dot{r}_{\rm II} = -\cII$ if the ionized gas is
effectively isothermal: recoil is just as important as pressure in
generating momentum.  The term in brackets on the right hand side of
(\ref{shellmom}) is therefore, to a good approximation, $(2, 1)\times
\cII^2$ for (blister, embedded) regions respectively, assuming a
D-critical ionization front for the former.

The rate at which mass is ionized is $\rhoII (\dot{r}_{\rm II} -
u_{\rm II}) A_{\rm sh}$, or approximately $\rhoII A_{\rm sh} |u_{\rm
II}|$ for a blister region when $\rII\gg R_{\rm St,0}$.  The ratio
between the rate of momentum generation and the rate of ionization is
$\phiII \cII$, so equation (\ref{shellmom}) implies 
\begin{equation}\label{phiIeval'd} 
\phiII = \frac{\cII}{-u_{\rm II}} + \frac{-u_{\rm II}}{\cII} \geq 2
\end{equation}
where equality holds for $u_{\rm II} = -\cII$, i.e., a D-critical
front. 

The expansion of the \HII region is simplest to determine when $\rII
\gg R_{\rm St, 0}$.  In this phase, $M_{\rm sh} \simeq A_{\rm sh}
\rho_0 \rII/3$ assuming radial expansion.  We seek a self-similar
expansion of the form $\rII\propto t^\eta$; equation (\ref{shellmom})
admits the solution\footnote{The solution for embedded regions was
presented in 1995 by C. F. McKee in lectures for Ay216 at
U.C. Berkeley.}
\begin{eqnarray}\label{rIIEval}
\rII &=& \left[\frac{(2, 1)\times 3}{\eta(4\eta-1)}\right]^{2/7} R_{\rm St,
0}^{3/7} (\cII t)^{4/7} \nonumber\\ &=&
(23,19) \times \left(\frac{t}{3.7~\Myr}\right)^{4/7}
\left(\frac{N_{H,22}}{1.5}\right)^{-3/7} \nonumber \\ &~& \times 
\Mclsix^{1/7} S_{49}^{1/7} ~\pc
\end{eqnarray}
for (blister, embedded) regions, respectively.  In this equation the first line
indicates $\eta=4/7$, which fixes the coefficient in the second line.
We have normalized to the ionization-averaged lifetime of rich OB
associations (M97) of 3.7 Myr.

The momentum of radial motion of the expanding shell is 
\begin{eqnarray}\label{MomentaBlEmb}
\delta p &=& (2.4,2.2)\times 10^5 \left(\frac{t}{3.7~\Myr}\right)^{9/7} 
\left(\frac{N_{H,22}}{1.5}\right)^{-3/14}\nonumber \\ &~& \times 
\Mclsix^{1/14}S_{49}^{4/7} ~\Msun ~\kps 
\end{eqnarray}
for (blister, embedded) regions.  Note that the extra thrust generated
at the ionization front in a blister region compensates for the
smaller working surface $A_{\rm sh}$.  Indeed, the two results are so
similar that we may estimate $\delta p$ using the intermediate
coefficient $2.3\times 10^5 ~\Msun~\kps$, without discriminating
between the two types of regions.  This will simplify the analysis
in \S \ref{PperStar}.

For a blister region, the mass evaporated is $\delta p/(\phiII
\cII)$: for $\phiII=2$,
\begin{eqnarray}\label{MevEval}
\delta M_{\rm dest} &=& 1.2 \times 10^{4}
\left(\frac{t}{3.7~\Myr}\right)^{9/7}
\left(\frac{N_{H,22}}{1.5}\right)^{-3/14} \nonumber \\ &~& \times 
\Mclsix^{1/14} S_{49}^{4/7} \Msun.
\end{eqnarray}
The axisymmetric calculations of \cite{1989A&A...216..207Y} give a
result that is only $6\%$ lower, once we account for the differences
between their ionized sound speed and recombination coefficient and
ours (without these corrections, their result would be $17\%$ higher).
This favorable comparison gives us confidence in the approximations
adopted in this section.  

Note that the above results differ quantitatively from those given by
\cite{1979MNRAS.186...59W}, which were adopted by
WM97 to study the erosion of molecular clouds
(see \S \ref{IFmomentum}).  Our equation (\ref{MevEval}) for the mass
evaporated agrees with these authors' results within $1\%$.  However,
equation (\ref{rIIEval}) gives a radius for a blister or an embedded
region that is smaller by a factor of 1.6 or 1.9, respectively, than
the characteristic size quoted by these authors for blister regions --
implying this characteristic size is best interpreted as the diameter
of the \HII region (note that $\delta p\propto \rho_0 \rII^4/t$). 

\subsubsection{Regime of Validity}\label{Validity}
The above equations must be restricted to \HII regions that are still
expanding and bounded within their GMC at the end of their ionizing
lifetimes.  If its luminosity is too low, the region will decelerate
to $\ceff$ and stall before its driving stars burn out; as WM97 argue,
such associations are too small to matter (but see \S \ref{Conclusions}).  If its luminosity is
too high, the \HII region will envelop its entire cloud ($\rII >
\Rcl)$, deforming the GMC into a cometary configuration
\citep{1990ApJ...354..529B} rocketing away from the association
\citep{1955ApJ...121....6O}.  Since this reduces the rate of
photoevaporation considerably, WM97 argued that an appropriate maximum
value for $\delta M_{\rm dest}$ should be roughly the value predicted for
a blister region at twice the time required for its size to match the
cloud radius.  Using \cite{1979MNRAS.186...59W}'s theory for the size
scale, WM97 identified a maximum value 
\begin{equation}\label{MevMax}  
\delta M_{\rm dest,max}\simeq 4.6 \times 10^4
\left(\frac{N_{H,22}}{1.5}\right)^{-3/8} \Mclsix^{7/8} S_{49}^{1/4}
\Msun.
\end{equation}  
Because the blister \HII region radius derived in equation
(\ref{rIIEval}) is smaller by a factor 1.6 than in
\cite{1979MNRAS.186...59W}'s theory, this upper limit corresponds to
evaluating equation (\ref{MevEval}) at a time when $\rII = 0.94\Rcl$.
Since this upper limit is quite uncertain, and since there is an
ambiguity between the radius and diameter of the \HII region in WM97's
argument, we shall simply adopt their value as given in equation
(\ref{MevMax}).  Correspondingly, we shall take an upper limit for the
effective injection of momentum to be 
\begin{equation}\label{pMax}
\delta p_{\rm max}\simeq 2\cII \delta M_{\rm dest,max}.
\end{equation}
Note that these upper limits on $\delta M_{\rm dest}$ and $\delta p$
are important for \HII regions if
\begin{equation}\label{S49MevMax}
S_{49} \gtrsim 63 \left(\frac{N_{H,22}}{1.5}\right)^{-3/8} \Mclsix^{5/2} \left(\frac{t_i}{3.7~\Myr}\right)^{-4}. 
\end{equation}

A more serious limitation on the theory presented here arises from the
fact that we have ignored gravity in the evolution of an \HII region. 
If the GMC is too dense ($n_H>140~\cm^{-3}$), its free-fall time will
be shorter than the typical ionizing lifetime of 3.7 Myr.  This is
typically true of clouds unless
\begin{equation}\label{MclLimitFromGrav}
\Mcl>3.7\times 10^5~\Msun. 
\end{equation}
For clouds below this limit, the orbital motion of an association
within or about the cloud is likely to alter its \HII region, possibly
by converting it into a cometary \HII region of the type discussed by
\cite{1969A&A.....1..431R}, \cite{1986ApJ...300..745R} and
\cite{1997RMxAA..33...73R}. 

Lastly, we have not attempted to account for the inhomogeneities of
molecular cloud material in the evolution of an \HII region
\citep[e.g.,][]{1995MNRAS.277..700D}, nor to instabilities that may
develop during its expansion \citep{1996ApJ...469..171G}.  The finite
porosity of the GMC (i.e., the interaction between \HII regions; see
WM97) will be accounted for in an approximate manner in \S
\ref{Porosity}.

\subsection{Contribution from Stellar Winds and Supernovae}
\label{Winds&SNe}
We now consider the effects of stellar winds and supernovae on the
evolution of an \HII region.  Results in this section are based on the
stellar evolution code Starburst99 \citep{1999ApJS..123....3L}, for
which we have used a Scalo stellar initial mass function normalized as
in MW97.  In the first 3.7 million years after stars form, they inject
roughly $78 M_\star ~\kps$ of momentum: $40\phi_w M_\star \kps$ from
protostellar winds, $30 M_\star \kps$ from main-sequence and evolved
stars, and $8 M_\star \kps$ in supernova ejecta.
In the formation of a sufficiently massive stellar cluster,
protostellar winds are mostly stopped within the self-gravitating
clump from which the cluster arose \citep{2000ApJ...545..364M} and
will affect this clump's dynamics \citep{BM96,1999sf99.proc..353M}
more than those of the surrounding cloud.  The remaining contribution is
\begin{equation}\label{pej}
\delta p_{\rm ej} \simeq 38 \left(\frac{t}{3.7~\Myr}\right) M_\star
\kps.
\end{equation}
If, in the first 3.7 Myr, this is smaller than the momentum imparted
by the \HII region, the ram pressure from stellar winds is lower than
$\rhoII\cII^2$.  This means that the winds will be confined within the
\HII region, while the ionizing stars shine, unless
\begin{equation}\label{WindRelevance}
S_{49} \gtrsim (270, 210) \times
\left(\frac{N_{H,22}}{1.5}\right)^{-1/2} \Mclsix^{1/6}
\end{equation}
for (blister, embedded) \HII regions, respectively.  Sufficiently
luminous associations couple primarily through stellar ejecta rather
than through their \HII regions; this is significant for the massive
clusters forming in starburst environments \citep{TanMcKee00} and for
the most luminous of Milky Way clusters.

Equation (\ref{WindRelevance}) accounts only for the momentum in
material flung away from a star.  More radial momentum can potentially
be generated in a pressurized bubble or blastwave that entrains
ambient mass, as the radial momentum varies with energy $E$ and total
mass $M$ as $\delta p \sim (E M)^{1/2}$.  Whereas in a thin shell
$\delta p$ is conserved while $E$ is not, in a fully adiabatic
bubble or blastwave $E$ is conserved while $\delta p$ increases as $M$
does.  Intermediate cases, such as pressure-driven snowplows
\citep{OM88} and partially radiative bubbles \citep{KM92a}, involve a
loss of energy but a gain of radial momentum.

\cite{1984ApJ...278L.115M} have argued that, in the context of an
inhomogeneous medium, stellar wind bubbles are either confined within
their \HII regions or made radiative by mass input from
photoevaporating clumps.  In a blister \HII region, furthermore, hot
gas can escape from the cloud.  For these reasons we shall assume that
stellar wind bubbles do not generate momentum significantly in excess
of the wind momentum itself.

Supernovae merit special attention, as supernova remnants typically
experience an adiabatic Sedov-Taylor phase ($\delta p\propto r^{3/2}$)
followed by a pressure-driven snowplow phase ($\delta p\propto
r^{1/2}$).  For an upper IMF cutoff of $120~\Msun$, the first
supernovae explode 3.6 Myr after the onset of star formation and
thereafter occur every $(3/S_{49})$ Myr for the first million years,
slowing to once per $(5/S_{49})$ Myr thereafter.  This frequency should
be compared with the sound-crossing time of the \HII region, $2
(\Mclsix S_{49})^{1/7} (N_{H,22}/1.5)^{-3/7}$ Myr, and with the
$e$-folding decay time of the ionizing luminosity, roughly 2.5 Myr.
In a rich association the delay between supernovae is the shortest of
these, implying that supernovae blend together with the stellar
winds. The above arguments indicate that the combined effects of
supernovae and winds are negligible unless equation
(\ref{WindRelevance}) is satisfied.

As a check, we have calculated the dynamics of the very first
supernova remnant inside the \HII region, using the theory of
\cite{1988ApJ...334..252C}.  Because the progenitor mass is very high
($\sim 90~\Msun$ in the presupernova state), the remnant becomes
radiative before a comparable mass has been swept up, skipping the
Sedov-Taylor phase \citep[see also][]{1980ApJ...237..781W}.  The
high progenitor mass also allows for a transition to a
momentum-conserving snowplow phase while the remnant is still
expanding supersonically.  The remnant becomes subsonic and merges
with the \HII region before striking its periphery if 
\begin{equation}\label{SNmerge}
S_{49} > (2.0, 2.7)\times  \left(\frac{N_{H,22}}{1.5}\right)^{0.66}
E_{51}^{1.2} \Mclsix^{-0.22}
\end{equation}
for (blister, embedded) regions.  If this happens, the remnant adds
little or no momentum to the \HII region.  This corroborates the above
conclusion that supernovae are not significant for the momentum of the
\HII region, except for regions smaller than the limit in equation
(\ref{SNmerge}) or larger than the limit in equation
(\ref{WindRelevance}).  The possibility remains that supernovae in
small associations contribute non-negligibly to the total momentum
input and hence reduce the ionizing flux and star formation rate we
derive below by neglecting supernovae; this merits further study. 

\subsection{Direct Disruption of Small Clouds?}\label{Oblit}
If a cluster were to deliver an impulse in excess of 
\begin{equation}\label{disruption}
\Mcl \vesc = 1.4\times 10^7 \left(\frac{N_{H,22}}{1.5}\right)^{1/4}
\Mclsix^{5/4} ~\Msun~\kps
\end{equation}
to its parent cloud in a time short compared to the cloud's dynamical
time, the cloud would be dynamically unbound.  This is an unattainably
large value for giant clouds with masses of $\sim 10^6~\Msun$, but
smaller clouds might be disrupted in this manner. The Taurus-Auriga
and Ophiucus clouds have $\Mclsix\simeq 0.01$, and may be
susceptible to disruption.  Unfortunately the likelihood and frequency
of this process is difficult to assess, as it depends on several
uncertain elements: $\delta p_{\rm max}$ (eq. [\ref{pMax}]), the
maximum size of a cluster that can form within a given cloud (WM97),
and the importance of gravity for \HII regions in small clouds (\S
\ref{Validity}).

\subsection{Turbulent Forcing Scale}\label{Forcing}
For turbulence driven by expanding shells, the relevant forcing scale
is the radius of a shell once it has decelerated and become subsonic
relative to $\ceff$.  This is
\begin{equation}\label{rmerge}
\rmerge \equiv \min\left\{\left[\frac{3 ~\delta p}{(2,4)\times \pi
\rho_0 \ceff}\right]^{1/3}, \Rcl\right\}
\end{equation}
for (blister, embedded) regions, respectively.  We do not allow
$\rmerge$ to exceed the cloud radius; ignoring stellar winds, this
limit becomes important when
\begin{equation}\label{rmergeIsRcl}
S_{49} > (54, 220) \times \left(\frac{\ceff}{0.57\vrms}\right)^{7/4}
\left(\frac{N_{H,22}}{1.5}\right)^{0.81} \Mclsix^{2.1}; 
\end{equation}
our estimate $\ceff \simeq 0.57 \vrms$ is based on \cite{Crete99}'s
formula for the mean pressure within GMCs. 

How is the effective forcing wavelength $\lambda_{\rm in}$ related to
the merging radius $\rmerge$ in equation (\ref{tdissNum})?
Diametrically opposed regions of a disturbance move in opposite
directions, and should be considered half a wavelength apart: this
suggests $\lambda_{\rm in} \simeq 4 \rmerge$.  A similar conclusion
follows from the fact that a shell's expansion could be considered a
quarter cycle of oscillation (and its collapse, a second quarter
cycle); for these reasons we set
\begin{equation}\label{lambdaAndRmerge}
\lambda_{\rm in} = 4 \phi_m r_{m}
\end{equation}
and consider $\phi_m$ an uncertain parameter of order unity.  We shall
find in \S\S \ref{SFRGalactic} and \ref{Porosity} that this is
roughly consistent with the Galactic ionizing luminosity and star formation rate.  Conversely, if
future numerical simulations indicate a value of the product
$\phi_m\phi$ much less than unity, then feedback from \HII regions
would not explain the ionizing luminosity and star formation rate in the inner Galaxy.  This
question should soon be addressed by simulations like those of
\cite{MacLow99outflows} and \cite{2001astro.ph..6509M}.  Note that
this requires $\ceff$ and $r_m$ to be identified in the simulation
volume, so that loss of energy by decelerating shells can be
discriminated from turbulent dissipation: combining equations
(\ref{equipart}), (\ref{tdissNum}), (\ref{regen}), and
(\ref{lambdaAndRmerge}), 
\begin{equation}\label{SuggestNumerics}
 \phi_m \phi= \frac{1}{2} \left(\frac{0.78 \Mcl
\vrms^3}{0.83 \dot{p}_{\rm in} r_m\ceff}\right)
\end{equation}
where $\dot{p}_{\rm in}$ refers to the creation of momentum by
explosive events in a simulation driven by point sources, and the
coefficients 0.78 and 0.83 should be adjusted to agree with the
dissipation rate and degree of equipartition, respectively, in
simulations of homogeneously driven turbulence (as in eqs.
[\ref{equipart}] and [\ref{tdissNum}]) that share global parameters
like the ratio of magnetic to gas pressure.  The numerical
determination of the product $\phi_m\phi$ is essential to evaluate the
theory presented here, or indeed to estimate stellar feedback
in general.

\section{Feedback from a Population of OB Associations}\label{PperStar}

We now wish to account for the dynamical feedback from the entire
population of \HII regions that will exist within a given GMC.  A
similar project was undertaken by MW97, who modeled the luminosity
function of \HII regions in the Galaxy, and WM97, who calculated the
lifetimes of GMCs after hypothesizing how the Galactic population of
\HII regions was distributed amongst GMCs.  Our discussion follows
these prior works to the greatest degree possible, except that we
shall solve self-consistently for the rate of star formation within a
given GMC.

MW97 adopted a \cite{1986FCPh...11....1S} stellar initial mass
function (IMF), with an upper cutoff at $120~\Msun$, normalized for 
a mean stellar mass $\mbarstar = 0.51~\Msun$.  They used the
results of \cite{1996ApJ...460..914V} for the ionizing fluxes and
lifetimes of massive stars: averaged over the IMF, this gives a mean
ionizing flux $\sbar = 8.9\times 10^{-4}$ and a mean ionizing lifetime
$\tmsbar = 3.7$ Myr.  They assumed that the fraction of OB
associations born with more than $\Nstar$ stars, which we denote
$F_a(>\Nstar)$, satisfies 
\begin{equation}\label{Faform}
\frac{dF_a}{d\ln \Nstar} \propto \frac{1}{\Nstar}, 
\end{equation}
from a lower limit of 100 up to a maximum of $5.5\times 10^5$
stars.  With these limits, MW97 were able to fit surveys of luminous
\HII regions \citep[e.g.,][]{1989ApJ...337..761K}, the total star
formation rate in the Galaxy, the birthrate of nearby associations,
and the Galactic recombination radiation.   

A cluster's ionizing luminosity is dominated by its most massive
members; this implies a distinction between {\em rich} clusters
\cite[$S_{49}\gtrsim 10$;][]{1989ApJ...337..761K}, which are populous
enough to sample the IMF up to the upper cutoff, and {\em poor}
clusters, which are not.  The ionizing luminosity of a poor cluster
reflects its most massive member, and rises only statistically with
$\Nstar$.  The cluster's ionizing lifetime likewise reflects this
star's main-sequence lifetime ($t_{\rm ms} \propto s_{49}^{-0.23}$;
\citealt{1996ApJ...460..914V}, MW97).  In contrast, a rich cluster's
ionizing luminosity and lifetime represent averages over the IMF:
$S_{49} = \sbar \Nstar$ and $t_i = \tmsbar$, respectively.  This
distinction introduces a turnover in the cluster luminosity function
$F_a(>S_{49})$ at the boundary between rich and poor clusters:
$dF_a/d\ln S_{49}\propto S_{49}^{-1}$ for rich clusters, but
statistical fluctuations cause a flattening of $F_a(>S_{49})$ for poor
clusters (\citealt{1989ApJ...337..761K} and MW97).  Associations near
this turnover dominate the feedback from \HII regions on GMCs, because
$\delta p$ (and for blister regions, $\delta M_{\rm dest}$) scale as
$S_{49}^{4/7}t_i^{9/7}$.  For instance, WM97 found that half of all
photoevaporation in clouds with $\Mclsix>0.1$ is accomplished by
regions with $S_{49}<3.7$.

It will be useful to define the cluster-weighted mean of a quantity $x$ as
\begin{equation}\label{def:<X>_a}
\left<x\right>_a
\equiv \int x ~{dF_a}.
\end{equation}
For MW97's Galactic \HII region luminosity function the mean mass per
association is $\left<M_\star\right>_a = 440 ~\Msun$, corresponding to
$\left<S_{49}\right>_a = 0.77$.  Other useful averages are listed in
table \ref{Moments}. 
\begin{deluxetable}{ccccc}
%\tablewidth{250pt}
\tablecaption{Moments of the Galactic luminosity function of \HII
regions \label{Moments}}
\tablehead{%
%------------------------------------------------------------------------
\colhead{Variable}&\colhead{j} &\colhead{k} &\colhead{$\left<S_{49}^j
(t_i/3.7~\Myr)^k\right>_a$} &\colhead{Eqn.} }
%------------------------------------------------------------------------
\startdata
$\left<\delta p\right>_a, \left<\delta M_{\rm dest}\right>_a$ &
4/7   & 9/7  & 0.499 & (\ref{VcharIIestimate}), (\ref{MdestIIestimate}) \\
$\dot{\cal N}_a, \dot{M}_\star$ &
16/21 & 12/7 & 0.572 & (\ref{Ndotassocs}), (\ref{MdotstarGMC}) \\
$t_{d0}$ & 
1/14  & 9/7  & 1.856 & (\ref{Lifetime}) \\
$Q_0$  &
3/7   & 19/7 & 2.128 & (\ref{Q0}) \\
\enddata
%------------------------------------------------------------------------
\tablecomments{Averages over the Galactic initial luminosity function
of OB associations, derived from a \cite{1986FCPh...11....1S} IMF and
a Monte-Carlo simulation as described by MW97.}
\end{deluxetable}

The Galactic population of GMCs, $\Ncl(>\Mcl)$, satisfies
$d\Ncl/d\ln\Mcl \propto \Mcl^{-\alpha}$ where $\alpha\simeq 0.6$ from
an undetermined lower limit to an upper limit of $6\times
10^6~\Msun$ (MW97).  MW97 argue that a given cloud cannot make
arbitrarily large OB associations. Taking the maximum cluster mass to
be $10\%$ of the cloud mass, they derive the upper limit
\begin{equation}\label{Su(Mcl)}
S_{49} < S_{u,49}(\Mcl) \equiv \min(490,172\Mclsix), 
\end{equation} 
which we also adopt.  \HII regions must therefore occur within a given
GMC in different proportions than they are found in the galaxy.  WM97
construct a cloud's luminosity function $F_{a,M}(>S_{49})$ by
assuming: 1. no clusters form above $S_{u,49}(\Mcl)$; 2. since GMCs
give birth to \HII regions, the Galactic luminosity function must
equal the sum of all GMCs' luminosity functions; and 3. within the
clouds that can form a given size of cluster, the birthrate of those
clusters is proportional to cloud mass.  These assumptions led to
MW97's equation (23) for the population of associations forming within
a given cloud.  Below, we shall solve for the star formation rate in a
given cloud under the hypothesis that it is supported by \HII regions
forming within it; this requires that assumption (3) be dropped.  If
the star formation rate within clouds scales as $\dot{M}_\star \propto
\Mcl^{\beta}$, a derivation analogous to MW97's equation (16) gives
\begin{equation}\label{F_a,m} 
F_{a,M}(>{\cal N}_\star) \propto \frac{H[{\cal
N}_{\star,u}(\Mcl)-{\cal N}_\star]} {1 - ({\cal N}_\star/1.2\times
10^6)^{\beta-\alpha}} F_a(>{\cal N}_\star), 
\end{equation}
where $H(x)=(1,0)$ for $(x>0,x<0)$ is the step function and ${\cal
N}_{\star,u}(\Mcl)\equiv S_{u,49}(\Mcl)/\sbar$.  MW97 assumed
$\beta=1$; below we derive $\beta=1.32$ for the mass range of
interest, and in \S \ref{Porosity} we find $\beta=1.38$ after
accounting for the interaction between \HII regions. % is accounted for.
The above birthrate distribution is used in Monte Carlo simulations
(as described by MW97) to produce the figures in this section.
However, the following argument indicates that the difference between
$F_a$ and $F_{a,M}$ is unimportant for clouds in the mass range where
our theory is valid.

\subsection{The Giant-Cloud Approximation}\label{GiantCloudAppx}
For analytical estimates it is useful to note that the associations
responsible for most of the mass ejection and energy injection
($S_{49}\sim 3.7$) are much smaller, for giant clouds, than those
associations whose existence or behavior are affected by their sizes
or the finite sizes of their parent clouds.  Specifically, we noted in
\S \ref{Validity} that our neglect of cloud gravity is only valid for
clouds with $\Mclsix>0.37$.  For such clouds, $S_{u,49}>64$; the
\HII regions whose maximum sizes (eq. [\ref{S49MevMax}]) and merging
radii (eq. [\ref{rmergeIsRcl}]) approach $\Rcl$ have $S_{49}>5$,
increasing rapidly with $\Mcl$; and those for which winds and
supernovae are important (eq. [\ref{WindRelevance}]) have
$S_{49}>180$.  These facts have two implications: 1. A cloud's
luminosity function $F_{a,M}(>S_{49})$ only differs from the Galactic
form $F_a(>S_{49})$ for luminosities near $S_{u,49}(\Mcl)$, so we may
neglect this difference in analytical estimates; and 2. We may assume
that \HII regions are entirely contained within their parent clouds.
These approximations, to which we refer collectively as the {\em
giant-cloud approximation}, are the same that led to MW97's equation
(40).

\begin{figure*}
\centerline{\epsfig{figure=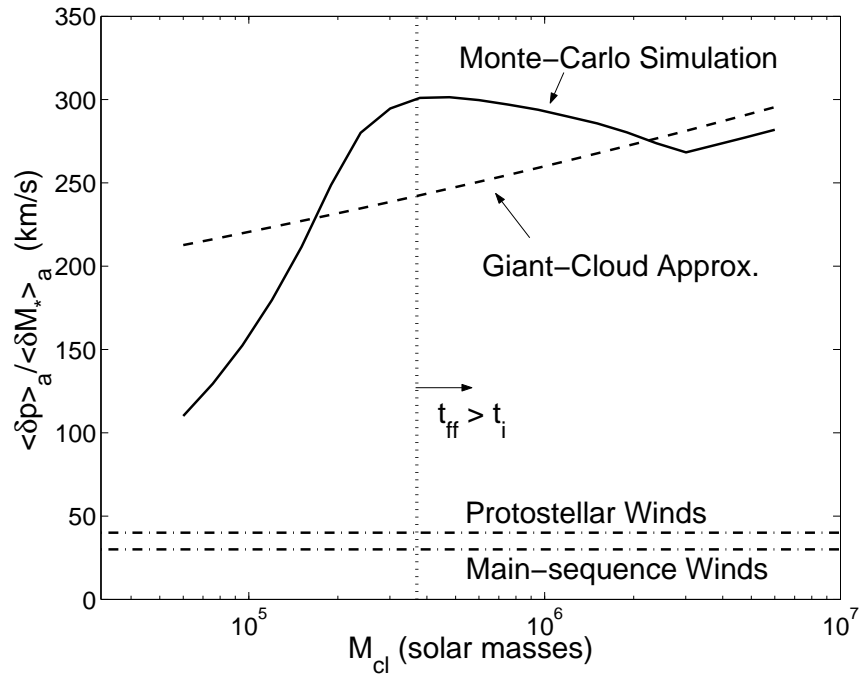 ,width=4.5in}} %,height = 3in, angle=0}}
\caption[Momentum per mass, and the giant-cloud approx.]
{\footnotesize Cloud momentum generated by \HII regions per stellar
mass formed, and an illustration of the giant-cloud approximation.
The theory for \HII regions presented here does not account for
gravity and is therefore only valid for clouds to the right of the
{\em vertical dotted line} -- those whose free-fall times are longer
than the ionizing lifetimes of OB associations
(eq. [\ref{MclLimitFromGrav}]).  The {\em solid line} is the result of
Monte-Carlo simulations that account for the difference between the
Galactic \HII region luminosity function and what is expected within a
given cloud (eq. [\ref{F_a,m}]; see also MW97).  These simulations
also account for the effect of finite cloud size.  The {\em dashed
line} represents the giant-cloud approximation in which both the
corrections to the Galactic luminosity function and the finite-size
effects are ignored.  To the right of the dotted line, differences
between the two curves result primarily from the luminosity function
(and are relatively minor): for instance, the kink at $\Mcl=3\times
10^6~\Msun$ divides clouds that can harbor the largest Galactic OB
associations from those that cannot.  Finite cloud-size effects
dominate to the left of the dotted line, where our model of \HII
regions is already invalidated by gravity.  The {\em dash-dot lines}
show the momenta of protostellar and main-sequence winds in the first
3.7 Myr of a cluster's life.
\label{fig0} }
\end{figure*}

For clouds with $\Mclsix<3.7$ it might be more appropriate to take
a small-cloud approximation in which all \HII regions are assumed
to outgrow their clouds: $\delta M_{\rm dest} = \delta M_{\rm dest,
max}$, $\delta p=\delta p_{\rm max}$, and $r_m = \Rcl$ (cf. MW97
eq. [41]).  This would be very uncertain, however, both because our
estimates of $\delta M_{\rm dest, max}$ and $\delta p_{\rm max}$ are
uncertain and also because we have neglected cloud gravity, which
should be important when $\Mcl<3.7\times 10^5~\Msun$. 

The performance of the giant-cloud approximation is illustrated in
figure \ref{fig0}, where we plot its prediction for $\left<\delta
p\right>_a$ ({\em dashed line}, eq. [\ref{MomentaBlEmb}]) against the
results of a Monte-Carlo simulation ({\em solid line}).  The
simulation accounts for the difference between the luminosity
distribution of OB associations within a given GMC ($F_{a,M}$,
eq. [\ref{F_a,m}]) and that within the Galaxy as a whole ($F_a$,
eq. [\ref{Faform}]), whereas this distinction is neglected in the
giant-cloud approximation.  This is the primary cause for the
difference between the two curves for $\Mcl > 3.7\times 10^5~\Msun$,
to the right of the {\em vertical dotted line}; for instance, the kink
in the simulation at $\Mclsix=3$ is caused by the kink in $F_{a,M}$ in
equation (\ref{F_a,m}).  The simulation also accounts for finite-size
effects ($\delta p_{\rm max}$, eq. [\ref{pMax}]); these dominate the
difference between the curves to the left of the dotted line.  However,
our neglect of cloud gravity is not valid in that region
%invalid to the left of the dotted line 
(eq. [\ref{MclLimitFromGrav}]). 

\subsection{Momentum Generation}\label{PopulationMomentum}
A comparison between $\left<\delta p\right>_a$ and
$\left<M_\star\right>_a$ gives the effective input of momentum per
stellar mass averaged over \HII regions.  In the giant-cloud
approximation, an integral of equation (\ref{MomentaBlEmb}) over
$F_a(>S_{49})$ gives
\begin{eqnarray}\label{VcharIIestimate} 
\left<\delta p\right>_a \simeq 260
\left(\frac{N_{H,22}}{1.5}\right)^{-3/14} \Mclsix^{1/14} ~\kps
\times \left<M_\star\right>_a. 
\end{eqnarray}
The characteristic velocity (momentum per stellar mass) identified in
this equation is much larger than the corresponding coefficient $\delta
p_w/m_\star = 40 \phi_w \kps$ that we have estimated for protostellar
winds in equation (\ref{pw}), or the value $38~\kps$ estimated for
main-sequence and evolved star ejecta in equation (\ref{pej}).  Since
equation (\ref{VcharIIestimate}) applies to both blister-type and
embedded \HII regions, it implies that \HII regions are more important
than the combined effects of protostellar winds, stellar winds, and
supernovae in driving turbulence within GMCs, regardless of the
efficiency of star formation.  Figure \ref{fig0} makes this point
graphically. 

Our estimate of $\delta p_{\rm dest,max}$ indicates that \HII regions
dominate over protostellar winds so long as $\Mcl\gtrsim 4\times
10^4~\Msun$.  This, in turn, suggests that protostellar winds may
support small clouds and the self-gravitating clumps within GMCs
\citep{BM96,1999sf99.proc..353M}, whereas \HII regions support the
GMCs themselves.

\subsection{Mass Ejection}\label{MevPerStar}
If we assume that all \HII regions evolve into blister regions, as
have \cite{1979MNRAS.186...59W}, \cite{1994ApJ...436..795F}, and
WM97, we arrive at an upper limit to the amount
of mass than can be removed from a GMC by photoevaporation.  Since
$\delta p = 2\cII \delta M_{\rm dest}$ for a blister region,
equation (\ref{VcharIIestimate}) gives, in the giant-cloud
approximation, 
\begin{equation}\label{MdestIIestimate}
\left<\delta M_{\rm dest}\right>_a \simeq 14
\left(\frac{N_{H,22}}{1.5}\right)^{-3/14} \Mclsix^{1/14} \times
\left<M_\star\right>_a, 
\end{equation}
implying a typical instantaneous star formation efficiency 
\begin{equation}\label{SFEtotestimate}
\varepsilon \equiv \frac{\left<M_\star\right>_a}{\left<\delta M_{\rm
dest}\right>_a + \left<M_\star\right>_a} \simeq 6.7 \% \times  
\left(\frac{N_{H,22}}{1.5}\right)^{3/14} \Mclsix^{-1/14} 
\end{equation}
\begin{figure*}
\centerline{\epsfig{figure=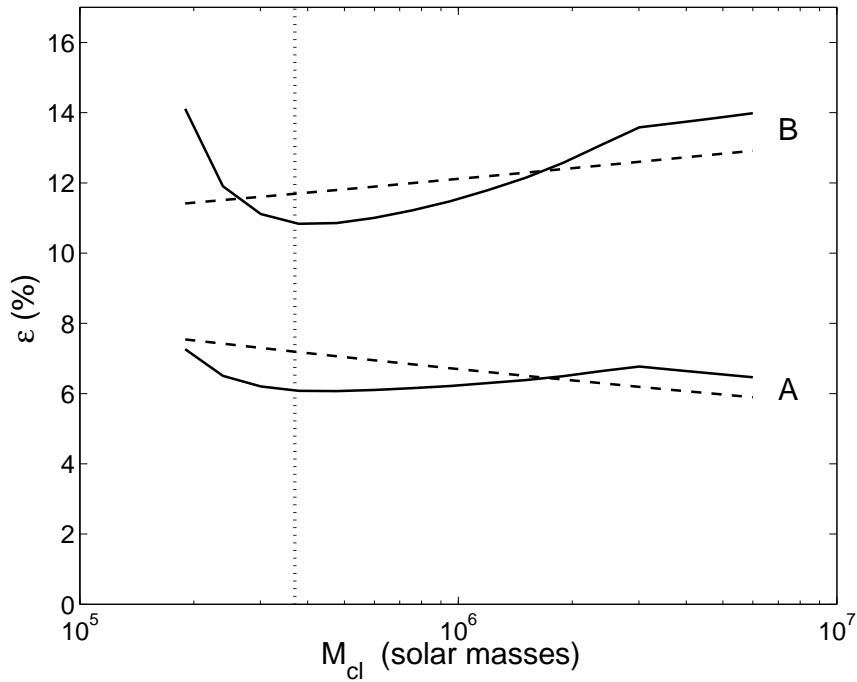 ,width=4.5in}} %,height = 3in, angle=0}}
\caption[Efficiency of star formation in giant clouds.] {\footnotesize
Star formation efficiency in giant clouds, assuming photoevaporation
is the dominant disruption mechanism.  The curves marked {\em A} are
not corrected for the interaction between \HII regions, whereas the
curves marked {\em B} account self-consistently for finite cloud
porosity (\S \ref{Porosity}).  The {\em solid lines} are Monte-Carlo
calculations based on MW97's model for the luminosity functions of OB
associations forming within a given cloud; effects due to finite cloud
size (eqs. [\ref{MevMax}], [\ref{pMax}], and [\ref{rmerge}]) are
included.  The {\em dashed lines} are analytical results of the
giant-cloud approximation (\S \ref{GiantCloudAppx}): equations
(\ref{SFEtotestimate}) and (\ref{SFEtotWithQ}) for cases {\em A} and
{\em B}, respectively.  In this and subsequent figures we assume that
all \HII regions enter the blister stage (see \S \ref{Porosity}).
\label{fig1} }
\end{figure*}
$\SFEtot$ and $\varepsilon$ are related by $\SFEtot =
\left<\varepsilon^{-1}\right>^{-1}$ \citep{2000ApJ...545..364M};
however, a sudden disruption of the cloud (a precipitous drop of
$\varepsilon$, as discussed in \S \ref{Oblit}) would reduce $\SFEtot$
below the typical value of $\varepsilon$. 

The lines marked {\em A} in figure \ref{fig1} compare equation
(\ref{SFEtotestimate}) with the results of a Monte-Carlo simulation
which accounts for the birthrates of OB associations within individual
clouds as given by equation (\ref{F_a,m}).  The
lines marked {\em B} account self-consistently for the frequent
interaction between \HII regions, in a manner we discuss in \S
\ref{Porosity}. 

Unfortunately, there are insufficient observations of clouds in the
mass range $3.7\times 10^5~\Msun<\Mcl<6\times 10^6~\Msun$ to test
equation (\ref{SFEtotestimate}) directly.  Observations of nearby
clouds, which are too small for an application of our theory, indicate
a significantly lower star formation efficiency than an extrapolation
of our theory would predict: for instance, observations of nearby
small clouds \citep{1991IAUS..147..293E} indicate that only $\sim 1\%$
of the mass is stellar and only $\sim 3\%$ will ever be.  Besides
gravity, a number of other effects may be important for clouds this
small; WM97 have emphasized their disruption by \HII regions that
outgrow their boundaries \citep[see also][]{1979ApJ...231..372E}, as
well as the likelihood that this disruption will render them
susceptible to photodissociation.

\section{Ionizing Luminosity and Star Formation Rate}\label{SFR}
As a cloud's turbulence decays it must be replenished, if virial
balance is to be maintained.  Energy can be derived from gravitational
contraction, winds and supernovae, and \HII regions.  Contraction is
problematic as it would imply a collapse rate close to free-fall,
hence rapid star formation; this may hold for the formation of stellar
associations, but cannot for entire GMCs \citep{1974ARA&A..12..279Z}.

We have shown that \HII regions are the most effective of the
remaining sources. Accordingly we will match the loss of turbulent
momentum with its regeneration in \HII regions.  We must account for a
population of \HII regions with different luminosities and therefore
different values of $\delta p$, $\rmerge$, and $\lambda_{\rm in}$.
Equations (\ref{diss}) and (\ref{tdissNum}) predict that energy decays
in a time proportional to $\lambda_{\rm in}$.  It is most consistent
to assume that the contribution from each type of \HII region decays
independently on its own timescale; the total turbulent energy is then
the sum of these decaying contributions.  Taking the decay time from
equation (\ref{tdissNum}), and taking the energy input $\delta E_{\rm
turb}(\delta p)$ from equation (\ref{regen}),
\begin{eqnarray}\label{Eturbdiss}
E_{\rm turb} &=& \sum_{S_{49}} \delta E_{\rm turb} \times ({\rm decay
~time})\times ({\rm rate})\nonumber \\ 
&=&\frac{\phi \ceff \dot{\cal N}_a}{2.4\vrms} \left<\lambda_{\rm
in}(\delta p)\delta p \right>_a, 
\end{eqnarray}
where $\lambda_{\rm in}(\delta p)$ is given by equations
(\ref{rmerge}) and (\ref{lambdaAndRmerge}) and $\dot{\cal N}_a$ is the
formation rate of associations.

Applied to a GMC, equation (\ref{Eturbdiss}) must be consistent with
the observed kinetic energy of the cloud and the expected degree of
equipartition between kinetic and magnetic energy (eq.
[\ref{equipart}]).  Equating expressions (\ref{equipart}) and
(\ref{Eturbdiss}) gives the formation rate of associations:
\begin{equation}\label{Ndotassocs}
\dot{\cal N}_a  = \frac{1.9 \Mcl \vrms^3}{\phi \ceff \left< \lambda_{\rm
in}(\delta p) \delta p \right>_a }. 
\end{equation}
If one were to vary the ionizing flux of all stars in the IMF by the
same factor, then $\dot{\cal N}_a$ would vary as $1/(\lambda_{\rm
in}\delta p)\propto (\delta p)^{-4/3} \propto
\left<S_{49}\right>_a^{-16/21} \propto \left<s_{49}\right>^{-16/21}$.
This scaling holds also for the star formation rate $\dot{M}_\star =
\left<M_\star\right>_a \dot{\cal N}_a$.  But, the mean ionizing flux
produced by these associations would vary much less: $S_{49,T}(\Mcl)
\equiv \left<S_{49}\right>_a \left<t\right>_{\rm ms} \dot{\cal N}_a
\propto \left<s_{49}\right>^{5/21}$. This is not surprising, as 
the ionizing photons are directly responsible for sustaining equilibrium. 

A cloud's ionizing luminosity is therefore quantity most tightly
constrained by the assumption of equilibrium: in the giant-cloud
approximation, 
\begin{eqnarray}\label{STotGMC}
S_{49,T}(\Mcl) &=& \frac{(37, 53)}{\phi_m(\phi/1.6)}
\left(\frac{N_{H,22}}{1.5}\right)^{1.37}
\left(\frac{\ceff}{0.57\vrms}\right)^{-2/3} \nonumber \\ &~& \times
\Mclsix^{37/28}. 
\end{eqnarray}
for \HII regions that are primarily in the (embedded, blister) state,
respectively.  This equation will be adjusted for the interaction
between \HII regions in \S \ref{Porosity}.  To compare 
% it or the updated form 
with observations of individual clouds one must account
for the leakage of ionizing photons; MW97 estimate that $\sim 70\%$ 
%of these
escape the observed \HII regions.

The star formation rate is $\dot{M}_\star = \left<M_\star\right>_a
\dot{\cal N}_a$.  Using the giant-cloud approximation and MW97's IMF,
\begin{eqnarray}\label{MdotstarGMC}
\dot{M}_\star 
&\simeq& \frac{(5.7,8.2)\times 10^{-3}}{\phi_m(\phi/1.6)}
\left(\frac{N_{H,22}}{1.5}\right)^{1.37}
\left(\frac{\ceff}{0.57\vrms}\right)^{-2/3} \nonumber \\ &~& \times
\Mclsix^{37/28} ~\Msun~{\rm yr}^{-1}
\end{eqnarray}
for (embedded, blister) regions.  Recall that $\dot{M}_\star \propto
\left<m_\star\right>\left<s_{49}\right>^{-16/21}$ is predicted less
robustly than $S_{49,T}(\Mcl)\propto \left<s_{49}\right>^{5/21}$ due
to uncertainties in the mean stellar mass and ionizing luminosity.
\cite{1994ApJ...435...22K} find that the ratio $\left<m_\star\right>/
(\left<s_{49}\right>\left<t_{\rm ms}\right>)$ varies by about $50\%$
among the forms of the IMF and sets of stellar tracks they consider;
the predicted $\dot{M}_\star$ is therefore uncertain by at least this
amount. 

The mean ionizing luminosity and lifetime depend on the metallicity
$Z$ of the stellar population, and these effects cause a shift in the
star formation rate relative to equation (\ref{MdotstarGMC}) even if
the IMF remains constant.  Numerical integrations of equation
(\ref{shellmom}), performed for Starburst99 populations of various
$Z$, exhibit $\left<\delta p\right>_a/ \left<M_\star\right>_a \propto
Z^{-0.092}$ implying $\dot{M}_\star \propto Z^{0.12} N_H^{1.37}$.
In any situation where M89's theory holds, $N_H\propto Z^{-1}$ and
therefore $\dot{M}_\star\propto Z^{-1.25}$.

\begin{figure*}
\centerline{\epsfig{figure=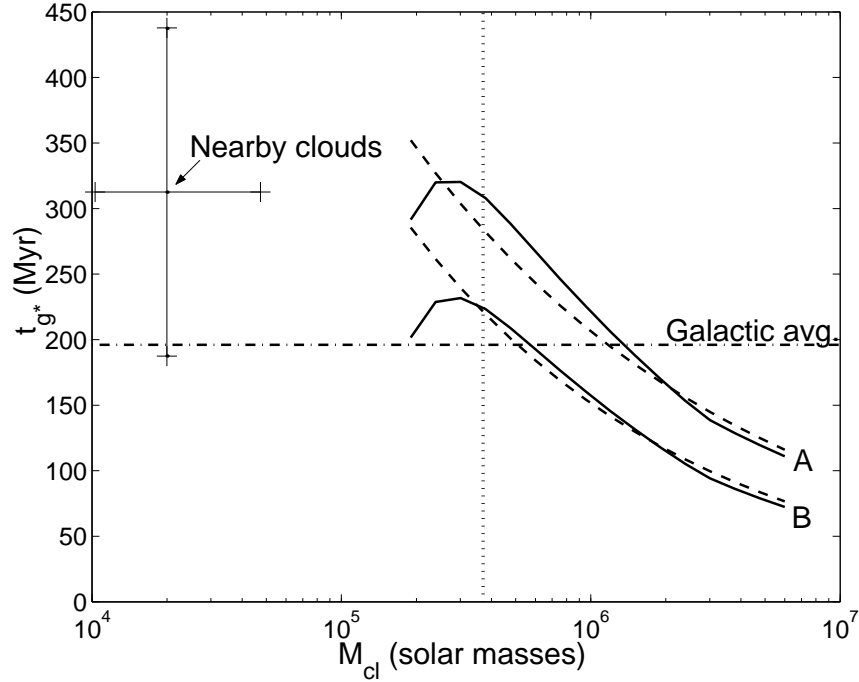,width=4.5in}} 
\caption[Star formation timescale in giant clouds.] {\footnotesize The
time scale for molecular gas to be converted into stars, $t_{g*}\equiv
\Mcl/\dot{M}_\star$.  Here, as in the other figures, the {\em vertical
dotted line} delimits the lower mass limit of validity of our theory
(eq. [\ref{MclLimitFromGrav}]); the {\em solid lines} are Monte-Carlo
calculations.  The {\em dashed lines} are the giant-cloud
approximation, equations (\ref{MdotstarGMC}) and (\ref{MdotstarWithQ})
for {\em A} and {\em B}, respectively.  The curves marked {\em A}
neglect interactions between \HII regions, whereas those marked {\em
B} account self-consistently for these interactions in the manner
derived in \S \ref{Porosity}.  The value of $t_{g*}$ for the entire
Galaxy is plotted as the {\em dash-dot line}; this corresponds to a
total molecular mass of $10^9~\Msun$ and star formation rate of
$5.1~\Msun$ per year \citep{1987sbge.proc....3M}.  Also plotted is the
value of $t_{g*}$ implied by \cite{2000AJ....120.3139C}'s
determination that the nearby Perseus, Orion A, Orion B, and Mon R2
clouds ($10^4~\Msun<\Mcl<3\times 10^4~\Msun$) have converted $\sim 1.6\%$ of
their mass into stars in the past 3-7 Myr.  In this figure the product
$\phi_m \phi$ is taken to be $1.6\times 1.2$. 
\label{fig2} }
\end{figure*}

Our hypothesis that the star formation rate is determined by a balance
between the dissipation of turbulence and its regeneration in \HII
regions can only make sense if it predicts more than one \HII region
per dynamical time of the cloud.  This is true for all the clouds
within the regime of validity of our theory, as we find that
$\dot{\cal N}_a \tff \gtrsim(13,19)\times
N_{H,22}^{13/21}(1.6/[\phi_m\phi])$ so long as $\Mclsix>0.37$.

In figure \ref{fig2} we plot $t_{g*} \equiv \Mcl/\dot{M}_\star$, the
time scale for gas to be converted into stars.  Equation
(\ref{MdotstarGMC}) is shown as the {\em dashed curve} marked {\em A}.
The {\em solid curve} marked {\em A} is the results of a Monte-Carlo
simulation which incorporates equation (\ref{Faform}) and the effects
of finite cloud size (eqs. [\ref{pMax}] and [\ref{rmerge}]). The
curves marked {\em B} are adjusted to account for the interaction
between \HII regions, as described in \S \ref{Porosity}.  Also plotted
({\em dash-dot line}) is value of $t_{g*}$ appropriate to the entire
inner Galaxy, and the range suggested by \cite{2000AJ....120.3139C}'s
observations of clouds around $2\times 10^4~\Msun$, 190-440 Myr
(corresponding to the acceptable age range 3-7 Myr for the embedded
population; see below).

\subsection{Galactic Ionizing Luminosity and Star Formation Rate} \label{SFRGalactic}
Equations (\ref{STotGMC}) and (\ref{MdotstarGMC}) refer only to a
single molecular cloud.  Integrating over the GMC population for the
inner Galaxy, we arrive at the total inner-Galactic ionizing
luminosity $S_T$ and star formation rate $\dot{M}_{\star T}$.  WM97
model the GMC population of the inner galaxy as $dN_{\rm cl}/d\ln\Mcl
\simeq 63 (6/\Mclsix)^{-0.6}$; setting $N_{H,22} = 1.5$ as observed,
and considering {\em only} giant clouds with $\Mclsix>3.7$ for which
we are relatively confident of the star formation rate, we find
\begin{equation}\label{STGalaxy}
S_T  \simeq \frac{(3.0,4.3)\times 10^{53}}{\phi_m(\phi/1.6)}
\left(\frac{\ceff}{0.57\vrms}\right)^{-2/3} ~{\rm s}^{-1}
\end{equation}
and
\begin{equation}\label{MdotTGalaxy}
\dot{M}_{\star T} \simeq \frac{(4.6, 6.6)}{\phi_m(\phi/1.6)}
\left(\frac{\ceff}{0.57\vrms}\right)^{-2/3} ~\Msun ~{\rm yr}^{-1}. 
\end{equation}
These results would be $15\%$ higher if we were to extrapolate the
giant-cloud approximation below the lower limit of
$\Mclsix=0.37$.  As before, $S_T$ is much better constrained than
$\dot{M}_{\star T}$ by our hypothesis that \HII regions dominate
feedback. 

We may use observations to check that clouds with $\Mclsix<0.37$,
whose evolution we cannot address, do not dominate the Galactic star
formation rate or ionizing flux.  Consider
\cite{2000AJ....120.3139C}'s observations of the Perseus, Orion A,
Orion B, and Mon R2 clouds ($\Mcl \sim 1-3\times 10^4 ~\Msun$ for
each).  \citeauthor{2000AJ....120.3139C} estimated the completeness of
the 2MASS survey and the resulting stellar fraction in these clouds, a
slowly increasing function of the assumed age of the observed stellar
population.  \citeauthor{2000AJ....120.3139C} argues that that these
clouds have created a stellar mass equal to $\sim 1.6\%$ of their own
mass in a period between 3 and 7 Myr.\footnote{This estimate is
independent of whether this age is associated with the lifespan of
such clouds.}  Extrapolating this star formation rate per unit
molecular mass to all clouds with $\Mclsix<0.37$, using MW97's cloud
mass function, gives a range of 0.7 to 1.7 $\Msun$ per year. This
estimate should be considered very uncertain, as it assumes
$\dot{M}_\star\propto \Mcl$ for small clouds; however it indicates
that clouds in the mass range $\Mcl < 3.7\times 10^5~\Msun$, which
constitute one third of the molecular mass, most likely do not
dominate the Galactic star formation rate.

MW97 estimate the ionizing flux of the inner Galaxy at $2.6\times
10^{53}~{\rm s}^{-1}$ (assuming all ionizing photons are caught within the
disk) and derive a star formation rate of $4.0~\Msun~{\rm
yr}^{-1}$. [\cite{1987sbge.proc....3M} estimates the star formation
rate in the inner Galaxy to be $5.1~\Msun~{\rm yr}^{-1}$.]  Equations
(\ref{STGalaxy}) and (\ref{MdotTGalaxy}) are comparable to this value,
with $\phi\simeq 1.6$ as suggested by \cite{Crete99}, and with
$\phi_m\simeq 1.1$ (close to unity as suggested in \S \ref{Forcing}).
Given the approximations that led to equation (\ref{MdotTGalaxy}),
this degree of agreement is quite remarkable; it supports our
proposition that feedback from \HII regions regulates the rate of star
formation in giant molecular clouds.

\section{Lifetimes of GMCs}\label{Lifetimes}
We may combine the star formation rate given in equation
(\ref{MdotstarGMC}) with the upper limit for mass evaporation from
GMCs given in equation (\ref{MdestIIestimate}), to arrive at a lower
limit for the lifetime of GMCs assuming all \HII
regions evolve into blister regions.  In the giant-cloud
approximation, this gives 
\begin{eqnarray}\label{Lifetime}
t_{d0} \equiv \frac{\Mcl}{\dot{M}_{\rm dest}} &\simeq& 13
\phi_m \left(\frac{\phi}{1.6}\right)
\left(\frac{N_{H,22}}{1.5}\right)^{-1.15}\nonumber \\ &~&\times 
\left(\frac{\ceff}{0.57\vrms}\right)^{2/3}  \Mclsix^{-11/28}
~\Myr.
\end{eqnarray}
for $\Mcl>3.7\times 10^5~\Msun$.  A significant population of \HII
regions that remain embedded rather than evolving into blister regions
would extend this lifetime, but we argue in \S \ref{Porosity} that
such a population is not likely.  Another effect that extends cloud
lifetimes relative to $t_{d0}$ is the finite porosity of \HII regions
within their volumes (MW97), for which we account in \S
\ref{Porosity}.
\begin{figure*}
\centerline{\epsfig{figure=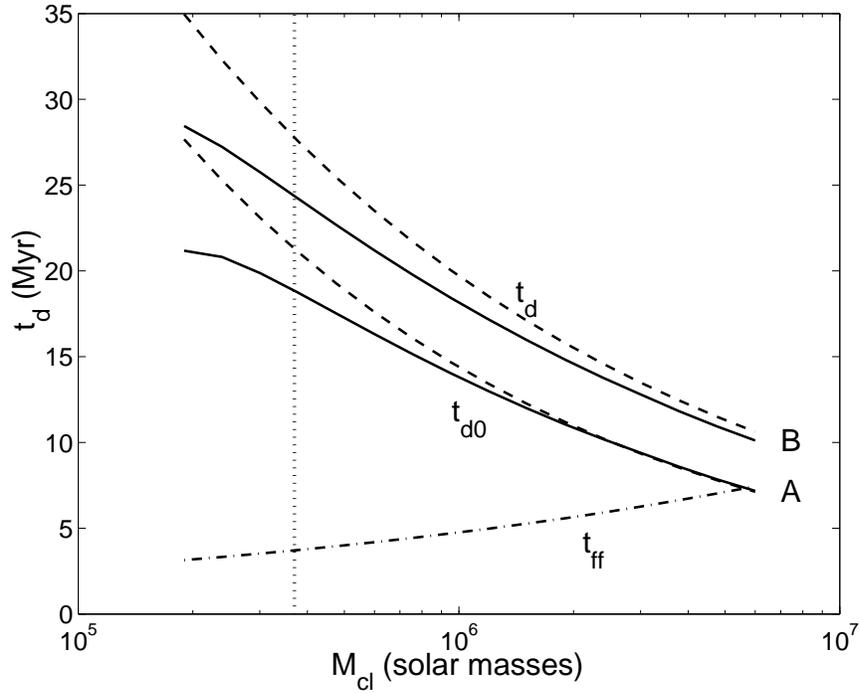 ,width=4.5in}} %,height = 3in, angle=0}}
\caption[Destruction time scale of giant clouds.] {\footnotesize
Destruction time scales $t_{d0}$ (case {\em A}, ignoring cloud
porosity) and $t_d$ (case {\em B}, accounting for finite porosity)
for giant clouds.  The {\em solid lines} are Monte-Carlo calculations;
the {\em dash-dot} lines, which result from the giant-cloud
approximation, represent equations (\ref{Lifetime}) and
(\ref{LifetimeWithQ}) for cases {\em A} and {\em B},
respectively.  Also plotted is the clouds' free-fall time $\tff$, which
approaches their destruction time at the upper mass limit for Galactic
GMCs. 
\label{fig3} }
\end{figure*}

Note that the estimate of the cloud destruction time in equation
(\ref{Lifetime}) is roughly $2.6 \Mclsix^{-4/7}$ times longer than the
cloud's free-fall time; the two time scales are equal for a mass $\Mcl
= 5.3 \times 10^6 (N_{H,22}/1.5)^{-13/12} ~\Msun$, very close to the
upper mass limit for Galactic GMCs ($\sim 6\times 10^6~\Msun$; WM97).
This raises the possibility that the upper mass limit derives from the
difficulty of assembling an object that destroys itself rapidly
compared to its free-fall time scale, a hypothesis that should be
tested against extragalactic observations.  We refine this argument in
\S \ref{Porosity}, where we adjust the destruction time for finite
cloud porosity.  (But note, a possible counterargument is raised in \S
\ref{Conclusions}.)

Cloud destruction timescales are plotted in figure \ref{fig3}, where
our Monte-Carlo simulation ({\em solid lines}) is compared with the
results of the giant-cloud approximation as given by equations
(\ref{Lifetime}) and (\ref{LifetimeWithQ}) for cases {\em A} and {\em
B}, respectively.  The curves marked {\em B} are adjusted for the
interaction between \HII regions (\S \ref{Porosity}).  The free-fall
time is also plotted, making visible the crisis of rapid destruction
for clouds above the Galactic upper mass limit. 

Our equation (\ref{Lifetime}) is based on the same physics that led
MW97 to estimate a destruction time of 20-25 Myr for clouds with
$\Mclsix> 0.1$.  Our estimate is shorter because MW97 assumed
$\dot{M}_\star\propto \Mcl$, whereas an energetic balance requires 
$\dot{M}_\star\propto \Mcl^{1.3}$ in our giant-cloud 
approximation; this redistribution shortens the lives of the massive
clouds to which our theory applies.

\section{Effect of Finite Porosity}\label{Porosity}
The frequency with which \HII regions form within GMCs implies that
they are likely to interact (WM97).  Defining the porosity $Q$ as the
time-averaged volume filling factor of \HII regions, 
$Q\equiv \dot{\cal N}_a \int \int^{t_{\rm ms}}
 [M_{\rm sh}(t)/\Mcl] dt\, dF_a$, 
we estimate from the above equations 
\begin{eqnarray}\label{Q0}
Q_0 &=& \frac{(2.6, 4.2)}{{\phi_m} (\phi/{1.6})}\times 
\left(\frac{N_{H,22}}{1.5}\right)^{19/12}\nonumber \\ &~&\times 
\left(\frac{\ceff}{0.57\vrms}\right)^{-2/3}  \Mclsix^{1/4}
\end{eqnarray}
for (blister, embedded) regions respectively.  This expression is not
yet fully self-consistent; we find $Q$ in terms of $Q_0$ below.
The large value of $Q_0$ for embedded regions implies that fully
embedded \HII regions are ruled out, as \HII regions will percolate
and allow their gas to vent.  This reinforces our neglect of embedded
regions in equation (\ref{Lifetime}). 
\begin{figure*}
\centerline{\epsfig{figure=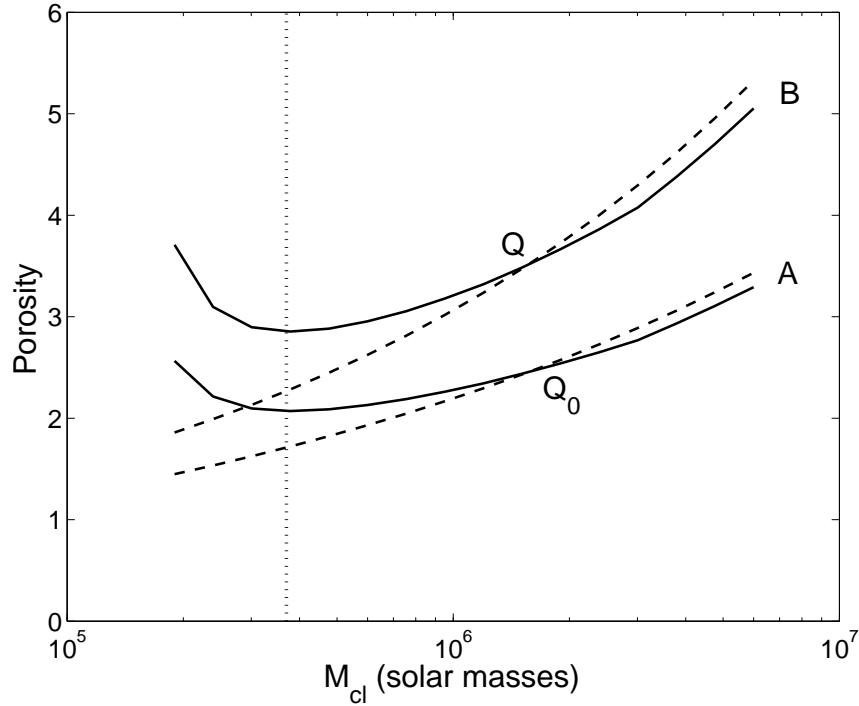 ,width=4.5in}} %,height = 3in, angle=0}}
\caption[Porosity of \HII regions in giant clouds.] {\footnotesize 
The porosity of \HII regions
within GMCs.  Plotted are the initial estimate ($Q_0$, case {\em A}),
which does not account self-consistently for the effect of porosity on
star formation rate, and the revised estimate ($Q$, case {\em B}),
which does.  The {\em dashed lines} result from the giant-cloud
approximation; case {\em A} is given by equation (\ref{Q0}). 
\label{fig4} }
\end{figure*}

WM97 argue that the effect of interactions is equivalent to regrouping
stars into non-interacting regions that are larger by the factor
$(1+Q)$, i.e., $\dot{\cal N}_a' = \dot{\cal N}_a/(1+Q)$ and $S_{49}' =
(1+Q)S_{49}$.  For each of the regrouped regions, $\delta p' =
(1+Q)^{4/7}\delta p$, $\delta M_{\rm dest}' = (1+Q)^{4/7}\delta M_{\rm
dest}$, and, for the giant-cloud approximation ($r_m<\Rcl$),
$\lambda_{\rm in}' = (1+Q)^{4/21}\lambda_{\rm in}$.  Whereas WM97
assumed a star formation rate within a given cloud, our theory demands
that we determine the star formation rate self-consistently by
matching turbulent decay with driving by \HII regions.  Applying these
transformations to equation (\ref{Ndotassocs}) and eliminating
$\dot{\cal N}_a'$, we find that $\dot{\cal N}_a =
(1+Q)^{5/21}\dot{\cal N}_{a0}$ where $\dot{\cal N}_{a0}$ is the
uncorrected value in equation (\ref{Ndotassocs}).  But, the porosity
is proportional the star formation rate: $Q = Q_0 \dot{\cal
N}_a/\dot{\cal N}_{a0}$.  A self-consistent value of $Q$ therefore
satisfies
\begin{equation}\label{Q}
{Q}{(1+Q)^{-5/21}}=Q_0, 
\end{equation}
so that $Q\simeq Q_0$ when $Q_0<1$, but $Q\simeq Q_0^{21/16}$ when
$Q_0 \gg 1$.  We find that the useful relation $(1+Q)= 1.85 Q_0$
holds within $1\%$ in the cloud mass range of interest.  The
analytical (equation [\ref{Q0}]) and numerical values of $Q_0$ and $Q$
are plotted in figure \ref{fig4}. 

To account self-consistently for finite porosity, we must increase
$\dot{\cal N}_a$, $S_{49,T}(\Mcl)$ and $\dot{M}_\star$ by
$(1+Q)^{5/21}$ relative to equations (\ref{Ndotassocs}),
(\ref{STotGMC}), and (\ref{MdotstarGMC}), increase the cloud lifetimes
by $(1+Q)^{4/21}$ relative to equation (\ref{Lifetime}), and increase
the stellar mass per ejected mass $(\varepsilon^{-1}-1)^{-1}\simeq
\varepsilon$ by $(1+Q)^{3/7}$ relative to equation
(\ref{SFEtotestimate}).  The revised equations are
\begin{eqnarray}\label{STotWithQ}
S_{49}(\Mcl)
&\simeq& \frac{54}{\left[\phi_m(\phi/1.6)\right]^{26/21}}
\left(\frac{N_{H,22}}{1.5}\right)^{1.75}
\left(\frac{\ceff}{0.57\vrms}\right)^{-0.83} \nonumber \\ &~& \times
\Mclsix^{1.38} ~\Msun~{\rm yr}^{-1},
\end{eqnarray}
\begin{eqnarray}\label{MdotstarWithQ}
\dot{M}_\star 
&\simeq& \frac{8.3\times 10^{-3}}{\left[\phi_m(\phi/1.6)\right]^{26/21}}
\left(\frac{N_{H,22}}{1.5}\right)^{1.75}
\left(\frac{\ceff}{0.57\vrms}\right)^{-0.83} \nonumber \\ &~& \times
\Mclsix^{1.38} ~\Msun~{\rm yr}^{-1},
\end{eqnarray}
\begin{eqnarray}\label{LifetimeWithQ}
t_{d} &\simeq& 17
\left[\phi_m \left(\frac{\phi}{1.6}\right)\right]^{0.81}
\left(\frac{N_{H,22}}{1.5}\right)^{-0.85}\nonumber \\ &~&\times 
\left(\frac{\ceff}{0.57\vrms}\right)^{0.54}  \Mclsix^{-0.35}
~\Myr, 
\end{eqnarray}
and 
\begin{eqnarray}\label{SFEtotWithQ}
\varepsilon  &\simeq& {13} \% \times  
{\left[\phi_m\left(\frac{\phi}{1.6}\right)\right]^{-3/7} }
\left(\frac{N_{H,22}}{1.5}\right)^{0.89} \nonumber
\\ &~&\times \left(\frac{\ceff}{0.57\vrms}\right)^{-2/7}
\Mclsix^{1/28},  
\end{eqnarray}
respectively.  The corrected lifetime is $1.2 (\phi_m\phi/1.6)^{0.81}
(\Mclsix/6)^{-0.6} \tff$, i.e., just over one free-fall time at
the upper mass limit for Galactic GMCs.  The total ionizing luminosity
and star formation rate
in the inner Galaxy due to clouds in the mass range $3.7\times
10^5~\Msun<\Mcl<6\times 10^6~\Msun$ become $4.5\times 10^{53}
(\phi_m\phi/1.6)^{-26/21}~{\rm s}^{-1}$ and 
$7.0 (\phi_m\phi/1.6)^{-26/21}~\Msun~\yr^{-1}$, respectively. These are consistent
with the observed rate if $\phi\simeq 1.6$ and $\phi_m\simeq 1.6$. 

Thus, our estimates of the Galaxy's ionizing luminosity and star
formation rate (with $\phi_m\phi\simeq 1\times 1.6$) are comparable to
the values adopted by MW97 if cloud porosity is not accounted for
(consistency requiring $\phi_m\phi \simeq 1.1\times 1.6 $) and
somewhat higher (consistency requiring $\phi_m \phi \simeq 1.6\times
1.6$) after the porosity correction is applied.  This level of
agreement suggests \HII regions are indeed responsible for maintaining
energetic equilibrium within GMCs.  What might explain the remaining
discrepancy?  First, numerical estimates of turbulent decay rates may
decrease as studies improve in resolution and less diffusive codes are
used, or as turbulent anisotropy is included
\citep{2001astro.ph..5235C}.  Second, the Galaxy's ionizing luminosity
could be higher than MW97 found, if a significant fraction of ionizing
photons escape the Galactic disk \citep[unlikely, but
controversial;][]{2001astro.ph.10043B} -- but conversely, additional
ionization from sources like supernova remnants
\citep{2000ApJ...541..218S} would increase the discrepancy.  Third,
the discrepancy results from our method of accounting for finite
porosity, a crude approximation when $Q>1$.  Fourth, we were not
entirely able to exclude a contribution of momentum from supernovae in
\S \ref{Winds&SNe}, especially for small associations.

Lastly, we have followed MW97 in approximating clusters' ionizing
luminosity as a step function of duration $\left<t_{\rm
ms}\right>=3.7$ Myr.  In reality, a numerical integration of equation
(\ref{shellmom}) using the ionization history predicted by a
Starburst99 synthesis indicates that $\delta p$ continues to grow
(albeit more slowly), gaining another $50\%$ after 6.6 Myr.  This
should effectively increase $\left<t_{\rm ms}\right>$ in the largest
clouds, whose free-fall times are long enough to permit this
expansion.  All of these topics merit further study.

\section{Conclusions}\label{Conclusions}

The main results of this paper are as follows. 
\begin{enumerate}
\item \HII regions are the most plentiful sources of energy for
the turbulence within giant molecular clouds; they are more
significant than the combined effects of protostellar winds,
main-sequence and evolved-star winds, and supernovae.  This result was
indicated by the low efficiency of star formation in GMCs in \S
\ref{Dominance} and demonstrated on more general grounds in \S\S
\ref{Winds&SNe} and \ref{PopulationMomentum}.
\item The input of turbulent energy by \HII regions occurs on scales
comparable to, but somewhat smaller than, the cloud radius.
Large-scale forcing minimizes the rate of turbulent decay, as
recently emphasized by \cite{2001ApJ...551..743B}. 
\item A balance between turbulent decay and the regeneration of
turbulence by \HII regions allows a prediction of the stellar ionizing
luminosity and (less robustly) the star formation rate.  We present
these results for clouds in the mass range $3.7\times
10^5~\Msun<\Mcl<6\times 10^6~\Msun$, in which the stellar ionizing
lifetime is briefer than the free-fall time.  The results are roughly
consistent with the total ionizing luminosity and star formation rate
of the inner Galaxy, provided that future numerical simulations verify
our estimate of the coupling between momentum input and turbulent
energy (eq. [\ref{SuggestNumerics}]).  Our estimate of
the ionizing flux in the inner Galaxy is $\sim 70\%$ higher than the
observed value; we list in \S \ref{Porosity} several possible
resolutions of this discrepancy. 
\item Because \HII regions also evaporate their clouds, an energetic
balance also implies a rate of photoevaporation and a destruction time
scale for GMCs.  We derive a destruction time of 17 to 24 times 
$(\Mcl/10^6\Msun)^{-1/3}$ million years, somewhat shorter than
was found by WM97. 
\item The upper mass limit for Milky Way GMCs  most likely derives
from the difficulty of assembling an object that destroys itself in a
single crossing time.  However, less massive clouds (at least down to
$3.7\times 10^5~\Msun$) survive for many crossing times, produce many
\HII regions per crossing time, and are in both energetic and
dynamical equilibrium.
\item So long as there exists a minimum optical depth or column
density required for star formation, the vigorous energetic feedback by \HII
regions provides a mechanism for the maintenance of cloud column
densities near the critical value, and hence, for the GMC line
width-size relation.  Massive clouds therefore follow the scenario
proposed by {M89}, but with \HII regions rather than protostellar
winds as the primary agents of feedback.  This conclusion is not
assured for clouds less massive than $3.7\times 10^5~\Msun$, for which
cloud gravity must be considered in the dynamics of \HII regions. 
\end{enumerate}
The theory we have presented is robust, in the sense that it applies
to massive GMCs (containing the bulk of the molecular mass) and
derives from the flattening of the luminosity function between rich
and poor OB associations -- a product solely of the upper mass limit
for stars -- rather than the detailed birthrate distribution of OB
associations.  So long as a galaxy's molecular mass is concentrated in
the most massive clouds, and so long as the birthrate of stellar
associations drops with ionizing luminosity more steeply than
$d\dot{\cal N}_a/d\ln S_{49}\propto S_{49}^{-4/7}$, energetic
equilibrium within molecular clouds determines its ionizing flux. 
This assertion can be tested by extragalactic observations.
Starbursts may however be an exception to this rule if, 
as discussed below, their GMCs are not in equilibrium.

Other observational tests of the theory presented here include the
variation of ionizing luminosity and star formation rate (figure
\ref{fig2}) with cloud mass and the high porosity of HII regions
(figure \ref{fig4}) for massive GMCs.  The variations of these
quantities with mean cloud column density can potentially be tested by
observations of the SMC.  The star formation efficiency (figure
\ref{fig1}) and cloud lifetime (figure \ref{fig3}) will be more
difficult to verify.

The inhomogeneity of giant molecular clouds is the greatest source of
uncertainty in the present work.  The interaction of \HII regions --
one source of inhomogeneity -- was treated in an approximate manner in
\S \ref{Porosity}, but future work must treat the interaction of \HII
regions with a realistic background cloud.  

We have noted that clouds more massive than $6\times 10^6~\Msun$
should not form in the Milky Way because they would be disrupted by \HII
regions in the time needed to assemble them.  However, we must also
note that clouds more massive than about $3.7\times 10^6
(N_{H,22}/1.5)^{-1}~\Msun$ may be incapable of driving champagne flows
because their escape velocities exceed the exhaust velocity
$\phiII\cII=19.4~\kps$ of ionized gas.  This fact does not affect our
derivation of the star formation rate in such clouds, but it does call
into question our derivation of the lifetime for the most massive
clouds (and hence our suggestion for the origin of the upper mass
limit).  Further work will be needed to resolve this issue.

Similarly, clouds more massive than about $ 1.9 \times 10^7
(N_{H,22}/1.5)^{-1}~\Msun$ cannot have supersonic \HII regions, as their
effective sound speeds exceed $10~\kps$, the sound speed of ionized
gas.  Objects created in this state or pushed into it by an increase
in external pressure can neither be disrupted nor even supported by
photoionization, and must collapse as rapidly as their turbulence
decays; this is a recipe for the efficient production of massive star
clusters observed to occur in starburst galaxies.  For instance,
\cite{1991ApJ...366L...5S} find values of $N_{H,22}$ in the range
$10^3$ to $10^{4.5}$ for the {\em mean} central molecular gas in a number of
starbursts; this is sufficient to crush all GMCs in the mass range
$\Mcl>3.7\times 10^5 ~\Msun$. 

\acknowledgements It is a pleasure to thank Chris McKee, Jonathan
Williams, and the referee, Mordecai-Mark Mac Low, for substantial
comments that led to improvements of this paper.  I am also grateful
to Peter Goldreich, Rob Kennicutt, Peter Martin, Eve Ostriker, Nick
Scoville, and Jim Stone for insightful discussions and suggestions.
This research was supported by an NSERC fellowship.

%\nocite{*}
%\bibliographystyle{apj}  % {apalike}{plain}
%\bibliography{/home/matzner/Thesis/mybib}
%\bibliography{mybib}

\end{document}